\documentclass[twocolumn]{aastex63} 

\usepackage[caption = false]{subfig}
\usepackage{graphicx}

\usepackage{xspace}

\newcommand\kms{km~s$^{-1}$\xspace}

\newcommand\asec{$^{\prime\prime}$\xspace}
\newcommand\twCO{$^{12}$CO\xspace}
\newcommand\thCO{$^{13}$CO\xspace}
\newcommand\CeO{C$^{18}$O\xspace}
\newcommand\HtCO{H$_2$CO\xspace}
\newcommand\cCtHt{c-C$_3$H$_2$\xspace}

\begin{document}

\title{Early Planet Formation in Embedded Disks (eDisk) III: A first high-resolution view of sub-mm continuum and molecular line emission toward the Class 0 protostar L1527 IRS}

\correspondingauthor{Merel L.R. van 't Hoff}
\email{mervth@umich.edu}

\author[0000-0002-2555-9869]{Merel L.R. van 't Hoff}
\affil{Department of Astronomy, University of Michigan, 1085 S. University Ave., Ann Arbor, MI 48109-1107, USA}

\author[0000-0002-6195-0152]{John J. Tobin}
\affil{National Radio Astronomy Observatory, 520 Edgemont Rd., Charlottesville, VA 22903 USA} 

\author[0000-0002-7402-6487]{Zhi-Yun Li}
\affil{Department of Astronomy, University of Virginia, 530 McCormick Rd., Charlottesville, VA 22904, USA}

\author[0000-0003-0998-5064]{Nagayoshi Ohashi}
\affil{Academia Sinica Institute of Astronomy \& Astrophysics, 11F of Astronomy-Mathematics Building, AS/NTU, No.1, Sec. 4, Roosevelt Rd, Taipei 10617, Taiwan, R.O.C.}

\author[0000-0001-9133-8047]{Jes K. J{\o}rgensen}
\affil{Niels Bohr Institute, University of Copenhagen, {\O}ster Voldgade 5--7, 1350, Copenhagen K, Denmark}

\author[0000-0001-7233-4171]{Zhe-Yu Daniel Lin}  
\affil{Department of Astronomy, University of Virginia, 530 McCormick Rd., Charlottesville, VA 22904, USA}

\author[0000-0003-3283-6884]{Yuri Aikawa}
\affil{Department of Astronomy, Graduate School of Science, The University of Tokyo, 7-3-1 Hongo, Bunkyo-ku, Tokyo 113-0033, Japan}

\author[0000-0002-8238-7709]{Yusuke Aso}
\affiliation{Korea Astronomy and Space Science Institute, 776 Daedeok-daero, Yuseong-gu, Daejeon 34055, Republic of Korea} 

\author[0000-0003-4518-407X]{Itziar de Gregorio-Monsalvo}
\affil{European Southern Observatory, Alonso de Cordova 3107, Casilla 19, Vitacura, Santiago, Chile}

\author[0000-0001-5782-915X]{Sacha Gavino}
\affil{Niels Bohr Institute, University of Copenhagen, {\O}ster Voldgade 5--7, 1350, Copenhagen K, Denmark}

\author[0000-0002-9143-1433]{Ilseung Han}
\affil{Division of Astronomy and Space Science, University of Science and Technology, 217 Gajeong-ro, Yuseong-gu, Daejeon 34113, Republic of Korea}
\affil{Korea Astronomy and Space Science Institute, 776 Daedeok-daero, Yuseong-gu, Daejeon 34055, Republic of Korea}

\author[0000-0003-2777-5861]{Patrick M. Koch}
\affil{Academia Sinica Institute of Astronomy \& Astrophysics, 11F of Astronomy-Mathematics Building, AS/NTU, No.1, Sec. 4, Roosevelt Rd, Taipei 10617, Taiwan, R.O.C.}

\author[0000-0003-4022-4132]{Woojin Kwon}
\affil{Department of Earth Science Education, Seoul National University, 1 Gwanak-ro, Gwanak-gu, Seoul 08826, Republic of Korea}
\affil{SNU Astronomy Research Center, Seoul National University, 1 Gwanak-ro, Gwanak-gu, Seoul 08826, Republic of Korea}

\author[0000-0002-3179-6334]{Chang Won Lee}
\affil{Division of Astronomy and Space Science, University of Science and Technology, 217 Gajeong-ro, Yuseong-gu, Daejeon 34113, Republic of Korea}
\affil{Korea Astronomy and Space Science Institute, 776 Daedeok-daero, Yuseong-gu, Daejeon 34055, Republic of Korea}

\author[0000-0003-3119-2087]{Jeong-Eun Lee}
\affil{Department of Physics and Astronomy, Seoul National University, 1 Gwanak-ro, Gwanak-gu, Seoul 08826, Korea}

\author[0000-0002-4540-6587]{Leslie W. Looney}
\affil{Department of Astronomy, University of Illinois, 1002 West Green St, Urbana, IL 61801, USA}

\author[0000-0002-0244-6650]{Suchitra Narayanan}
\affil{Institute for Astronomy, University of Hawai‘i at Mānoa, 2680 Woodlawn Dr., Honolulu, HI 96822, USA}

\author[0000-0002-9912-5705]{Adele Plunkett}
\affil{National Radio Astronomy Observatory, 520 Edgemont Rd., Charlottesville, VA 22903 USA} 

\author[0000-0003-4361-5577]{Jinshi Sai (Insa Choi)}
\affil{Academia Sinica Institute of Astronomy \& Astrophysics, 11F of Astronomy-Mathematics Building, AS/NTU, No.1, Sec. 4, Roosevelt Rd, Taipei 10617, Taiwan, R.O.C.}

\author[0000-0001-6267-2820]{Alejandro Santamar{\'i}a-Miranda }
\affil{European Southern Observatory, Alonso de Cordova 3107, Casilla 19, Vitacura, Santiago, Chile}

\author[0000-0002-0549-544X]{Rajeeb Sharma}, 
\affil{Niels Bohr Institute, University of Copenhagen, {\O}ster Voldgade 5--7, 1350, Copenhagen K, Denmark}

\author[0000-0002-9209-8708]{Patrick D. Sheehan}
\affil{National Radio Astronomy Observatory, 520 Edgemont Rd., Charlottesville, VA 22903 USA} 

\author[0000-0003-0845-128X]{Shigehisa Takakuwa}
\affil{Department of Physics and Astronomy, Graduate School of Science and Engineering, Kagoshima University, 1-21-35 Korimoto, Kagoshima, Kagoshima 890-0065, Japan}
\affil{Academia Sinica Institute of Astronomy \& Astrophysics, 11F of Astronomy-Mathematics Building, AS/NTU, No.1, Sec. 4, Roosevelt Rd, Taipei 10617, Taiwan, R.O.C.}

\author[0000-0003-0334-1583]{Travis J. Thieme}
\affiliation{Institute of Astronomy, National Tsing Hua University, No. 101, Section 2, Kuang-Fu Road, Hsinchu 30013, Taiwan}
\affiliation{Center for Informatics and Computation in Astronomy, National Tsing Hua University, No. 101, Section 2, Kuang-Fu Road, Hsinchu 30013, Taiwan}
\affiliation{Department of Physics, National Tsing Hua University, No. 101, Section 2, Kuang-Fu Road, Hsinchu 30013, Taiwan}

\author[0000-0001-5058-695X]{Jonathan P. Williams}
\affil{Institute for Astronomy, University of Hawai‘i at Mānoa, 2680 Woodlawn Dr., Honolulu, HI 96822, USA}

\author[0000-0001-5522-486X]{Shih-Ping Lai}
\affiliation{Institute of Astronomy, National Tsing Hua University, No. 101, Section 2, Kuang-Fu Road, Hsinchu 30013, Taiwan}
\affiliation{Center for Informatics and Computation in Astronomy, National Tsing Hua University, No. 101, Section 2, Kuang-Fu Road, Hsinchu 30013, Taiwan}
\affiliation{Department of Physics, National Tsing Hua University, No. 101, Section 2, Kuang-Fu Road, Hsinchu 30013, Taiwan}
\affil{Academia Sinica Institute of Astronomy \& Astrophysics, 11F of Astronomy-Mathematics Building, AS/NTU, No.1, Sec. 4, Roosevelt Rd, Taipei 10617, Taiwan, R.O.C.}

\author[0000-0002-4372-5509]{Nguyen Thi Phuong}
\affiliation{Korea Astronomy and Space Science Institute, 776 Daedeok-daero, Yuseong-gu, Daejeon 34055, Republic of Korea\\}
\affiliation{Department of Astrophysics, Vietnam National Space Center, Vietnam Academy of Science and Techonology, 18 Hoang Quoc Viet, Cau Giay, Hanoi, Vietnam}

\author[0000-0003-1412-893X]{Hsi-Wei Yen}
\affil{Academia Sinica Institute of Astronomy \& Astrophysics, 11F of Astronomy-Mathematics Building, AS/NTU, No.1, Sec. 4, Roosevelt Rd, Taipei 10617, Taiwan, R.O.C.}




\begin{abstract}

\noindent Studying the physical and chemical conditions of young embedded disks is crucial to constrain the initial conditions for planet formation. Here, we present Atacama Large Millimeter/submillimeter Array (ALMA) observations of dust continuum at $\sim$0\farcs06 (8 au) resolution and molecular line emission at $\sim$0\farcs17 (24 au) resolution toward the Class 0 protostar L1527 IRS from the Large Program eDisk (Early Planet Formation in Embedded Disks). The continuum emission is smooth without substructures, but asymmetric along both the major and minor axes of the disk as previously observed. The detected lines of \twCO, \thCO, \CeO, \HtCO, \cCtHt, SO, SiO, and DCN trace different components of the protostellar system, with a disk wind potentially visible in \twCO. The \thCO brightness temperature and the \HtCO line ratio confirm that the disk is too warm for CO freeze out, with the snowline located at $\sim$350 au in the envelope. Both molecules show potential evidence of a temperature increase around the disk--envelope interface. SO seems to originate predominantly in UV-irradiated regions such as the disk surface and the outflow cavity walls rather than at the disk--envelope interface as previously suggested. Finally, the continuum asymmetry along the minor axis is consistent with the inclination derived from the large-scale (100\asec or 14,000 au) outflow, but opposite to that based on the molecular jet and envelope emission, suggesting a misalignment in the system. Overall, these results highlight the importance of observing multiple molecular species in multiple transitions to characterize the physical and chemical environment of young disks. \\
\end{abstract}



\section{Introduction} \label{sec:intro}

Planets form in disks around young stars, starting with the growth of dust grains beyond interstellar medium sizes. Evidence for planet formation already being underway when the disk is still embedded in its natal envelope has been inferred from low dust opacity spectral indexes in Class 0 sources \citep{Kwon2009,Shirley2011}, dust polarization \citep[e.g.,][]{Kataoka2015,Kataoka2016,Yang2016}, decreasing dust masses derived from (sub-)millimeter observations for more evolved systems \citep[e.g.,][]{Williams2019,Tychoniec2020}, and the lack of CO isotopologue emission toward the protostellar position due to grain growth in the Class~I system TMC1A \citep{Harsono2018}. In addition, rings in continuum emission, which could be a signpost of forming planets \citep[e.g.,][]{Bryden1999,Zhu2014,Dong2018}, are observed in disks as young as only $\sim$0.5 Myr \citep{ALMAPartnership2015,Segura-Cox2020,Sheehan2020}. Characterizing the physical and chemical conditions in young disks is thus crucial in understanding disk evolution and planet formation. 

L1527 IRS (also known as IRAS 04368+2557) is the first Class 0 source toward which a Keplerian rotating disk was established \citep{Tobin2012}. This low-mass protostar is located in the L1527 dark cloud in the Taurus star-forming region and has been observed extensively from the near-infrared to centimeter wavelengths. Based on Gaia Data Release 2 (DR2), \citet{Luhman2018} measured a distance of 139--141 pc for L1527, consistent with the analysis of Gaia DR2 and very long baseline interferometry  (VLBI) data by \citet{Galli2019}. \citet{Roccatagliata2020} group L1527 into the much larger Taurus B region with an average distance of 131.0 $\pm$ 1.0 pc, but sources in L1527 have parallaxes closer to the lower end of the range for Taurus B ($\sim$6.95 mas, corresponding to 143.9 pc). We therefore adopt a distance of 140 pc, which is also consistent with the distance used in earlier works by \citet{Kenyon1994} and \citet{Zucker2019}.

\begin{deluxetable*}{lccccccccc}
\tablecaption{Overview of molecular lines. \label{tab:molecularlines}}
\tablewidth{0pt}
\addtolength{\tabcolsep}{-1pt} 
\tabletypesize{\scriptsize}
\tablehead{
\colhead{Species} & \colhead{Transition} & \colhead{Frequency} & \colhead{$E_{\rm{up}}$ $^a$} & \colhead{$A_{ij}$ $^b$} & \colhead{$\Delta v$ $^c$} & \colhead{RMS $^d$} & \colhead{Velocity range $^e$} \\ [-0.25cm]
\colhead{} & \colhead{} & \colhead{(GHz)} & \colhead{(K)} & \colhead{(s$^{-1}$)} & \colhead{(\kms)} & \colhead{(mJy beam$^{-1}$)} & \colhead{(\kms)} \\ [-0.45cm]
} 
\startdata 
$^{12}$CO    & 2--1                      & 230.538000 & 16.6 & 6.910$\times10^{-7}$ & 0.635 & 0.95 & $-$11.30 -- $-$1.13, 0.77 -- 9.04 \\ 
$^{13}$CO    & 2--1                      & 220.398684 & 15.9 & 5.066$\times10^{-7}$ & 0.167 & 1.96 & $-$4.38 -- $-$0.21, 0.46 -- 3.97\\ 
C$^{18}$O    & 2--1                      & 219.560354 & 15.8 & 6.011$\times10^{-7}$ & 0.167 & 1.49 & $-$3.55 -- 3.64 \\
DCN          & 3--2                      & 217.238538 & 20.9 & 4.575$\times10^{-4}$ & 1.340 & 0.55 & $-$3.84 -- 1.53 $^g$\\
SO           & 6$_5$--5$_4$              & 219.949442 & 35.0 & 1.335$\times10^{-4}$ & 0.167 & 1.78 & $-$4.05 -- 3.30 \\
SiO          & 5--4                      & 217.104980 & 31.3 & 5.196$\times10^{-4}$ & 1.340 & 0.58 & $-$13.45 -- 1.30 \\
H$_2$CO      & 3$_{0,3}$--2$_{0,2}$      & 218.222192 & 21.0 & 2.818$\times10^{-4}$ & 1.340 & 0.51 & $-$3.84 -- 2.87 \\ 
H$_2$CO      & 3$_{2,1}$--2$_{2,0}$      & 218.760066 & 68.1 & 1.577$\times10^{-4}$ & 0.167 & 1.42 & $-$3.05 -- 3.14 \\
H$_2$CO      & 3$_{2,2}$--2$_{2,1}$      & 218.475632 & 68.1 & 1.571$\times10^{-4}$ & 1.340 & 0.48 & $-$2.50 -- 2.87 \\
c-C$_3$H$_2$ & 6$_{0,6}$--5$_{1,5}$ $^f$ & 217.822148 & 38.6 & 5.396$\times10^{-4}$ & 1.340 & 0.54 & $-$1.16 -- 1.53 \\ 
c-C$_3$H$_2$ & 6$_{1,6}$--5$_{0,5}$ $^f$ & 217.822148 & 38.6 & 5.396$\times10^{-4}$ & 1.340 & 0.54 & $-$1.16 -- 1.53 \\
c-C$_3$H$_2$ & 5$_{1,4}$--4$_{2,3}$      & 217.940046 & 35.4 & 4.026$\times10^{-4}$ & 1.340 & 0.52 & $-$1.16 -- 1.53 \\ 
c-C$_3$H$_2$ & 5$_{2,4}$--4$_{1,3}$      & 218.160456 & 35.4 & 4.041$\times10^{-4}$ & 1.340 & 0.50 & $-$1.16 -- 1.53 \\ 
CH$_3$OH     & 4$_2$--3$_1$, E & 218.440063 & 45.6 & 4.686$\times10^{-5}$ & 1.340 & 0.49 & --  \\
\enddata
\vspace{-0.2cm}
\tablenotetext{a}{Energy of the transition's upper level.}
\vspace{-0.3cm}
\tablenotetext{b}{Einstein A coefficient of the transition.}
\vspace{-0.3cm}
\tablenotetext{c}{Velocity resolution of the observations.}
\vspace{-0.3cm}
\tablenotetext{d}{RMS level per channel in the data cubes imaged with a robust parameter of 2.0, measured within a 10$^{\prime\prime}$ region over five empty channels.}
\vspace{-0.3cm}
\tablenotetext{e}{Velocity range over which emission is detected ($>3\sigma$) with the system velocity of 5.9 \kms shifted to 0 \kms \citep{Caselli2002a,Tobin2011}.}
\vspace{-0.3cm}
\tablenotetext{f}{These two transitions are blended.}
\vspace{-0.3cm}
\tablenotetext{g}{DCN is only detected at the 3--4$\sigma$ level.}
\end{deluxetable*}

L1527 IRS (hereafter L1527) is often classified as a borderline Class 0/I object, as classification is challenging due to the edge-on orientation. Its bolometric temperature and submillimeter luminosity to bolometric luminosity ratio are typical of a Class 0 source, but at a lower inclination it would be classified as a Class I object \citep{Tobin2008}. The large envelope mass and extended outflow cavities suggest that L1527 is younger than typical Class I sources, but it lacks the collimated outflow of typical Class 0 sources (see e.g., the discussion in \citealt{Tobin2013}). Recent re-analysis of the spectral energy distribution (SED) classified L1527 as a Class~0 source with a bolometric luminosity of 1.3 $L_\sun$ and a bolometric temperature of 41 K \citep{Ohashi2022_eDisk}. 

Single-dish sub-millimeter observations of L1527 have identified a bipolar outflow in $^{12}$CO emission with an orientation almost perfectly in the plane of the sky \citep{Tamura1996,Hogerheijde1998}. Bright bipolar scattered light nebulae extending $\sim$10,000 au along the east--west outflow axis are visible in infrared observations with the Spitzer Space Telescope and ground-based telescopes \citep{Tobin2008,Tobin2010}, as well as in the recently released JWST NIRCam image (release id 2022-055, PI: K. Pontoppidan\footnote{https://webbtelescope.org/contents/news-releases/2022/news-
2022-055}). The eastern outflow lobe harbors a compact ($\sim$1\asec long) radio continuum jet close to the protostellar position at centimeter wavelengths \citep{Reipurth2004}. 

\citet{Ohashi1997} identified a flattened infalling and rotating envelope with a radius of 2000 au from 6\asec resolution C$^{18}$O observations. The presence of a rotationally supported disk was initially inferred from $^{13}$CO observations with the Combined Array for Millimeter-wave Astronomy (CARMA; \citealt{Tobin2012}) that also resolved the continuum at 0\farcs35 resolution, and was later confirmed by observations with the Atacama Large Millimeter/sub-millimeter Array (ALMA; \citealt{Ohashi2014,Aso2017}). High-resolution (0\farcs15) continuum observations with ALMA have suggested that the disk is warped with the inner and outer disk boundary between 40 and 60 au \citep{Sakai2019}. Even higher resolution (0\farcs08) observations with the Karl G. Jansky Very Large Array (VLA) initially revealed clumpy substructures in $Q$-band \citep{Nakatani2020}, but these structures were not confirmed in later observations with higher sensitivity \citep{Sheehan2022}.

Molecular line observations have shown that the disk is warm ($\gtrsim$ 20 K), based on the presence of CO gas out to at least 75 au \citep{vantHoff2018b}. The water snowline is suggested to be located between $\sim$2--4 au \citep{vantHoff2022}, but only a tentative detection of a complex molecule (methanol, CH$_3$OH) in the inner disk has been reported \citep{Sakai2014b,vantHoff2020}. Observations by \citet{Sakai2014b,Sakai2014a} reveal different morphologies and kinematics for several molecular species, suggesting that they trace different components of the protostellar system. In particular, SO seems to be enhanced in a ring at the disk-envelope interface (see also \citealt{Ohashi2014}). 

While L1527 has been studied at high resolution at multiple wavelengths in continuum emission (0\farcs045 at cm wavelengths and 0\farcs08 at mm wavelengths; \citealt{Sheehan2022,Ohashi2022}, respectively), molecular line observations have been limited to a resolution of $\sim$0.3--0\farcs5. Here, we present high angular resolution 1.3 mm continuum (0\farcs06) and molecular line images (0\farcs17) obtained with the ALMA Large Program eDisk (Early Planet Formation in Embedded Disks). The molecular lines detected toward L1527 are \twCO, \thCO, \CeO, \HtCO, \cCtHt, SO, SiO and DCN. A notable nondetection is CH$_3$OH. The observations are described in Sect.~\ref{sec:Observations}. The structure of the 1.3 mm continuum is presented in Sect.~\ref{sec:Continuum} and the morphology and spatial origin of the molecular lines in Sect.~\ref{sec:Lines}. The system's inclination is discussed in Sect.~\ref{sec:Inclination}, a dynamical estimate of the central mass is made in Sect.~\ref{sec:DynamicalMass}, and the temperature structure is described in Sect.~\ref{sec:Temperature}. In Sect.~\ref{sec:MolecularStructure} we discuss the physical and chemical reasons behind the different molecular distributions. Finally, our conclusions are summarized in Sect.~\ref{sec:Conclusions}.

\begin{figure*}
\centering
\includegraphics[width=\textwidth,trim={0.5cm 8.2cm 0.5cm 1.0cm},clip]{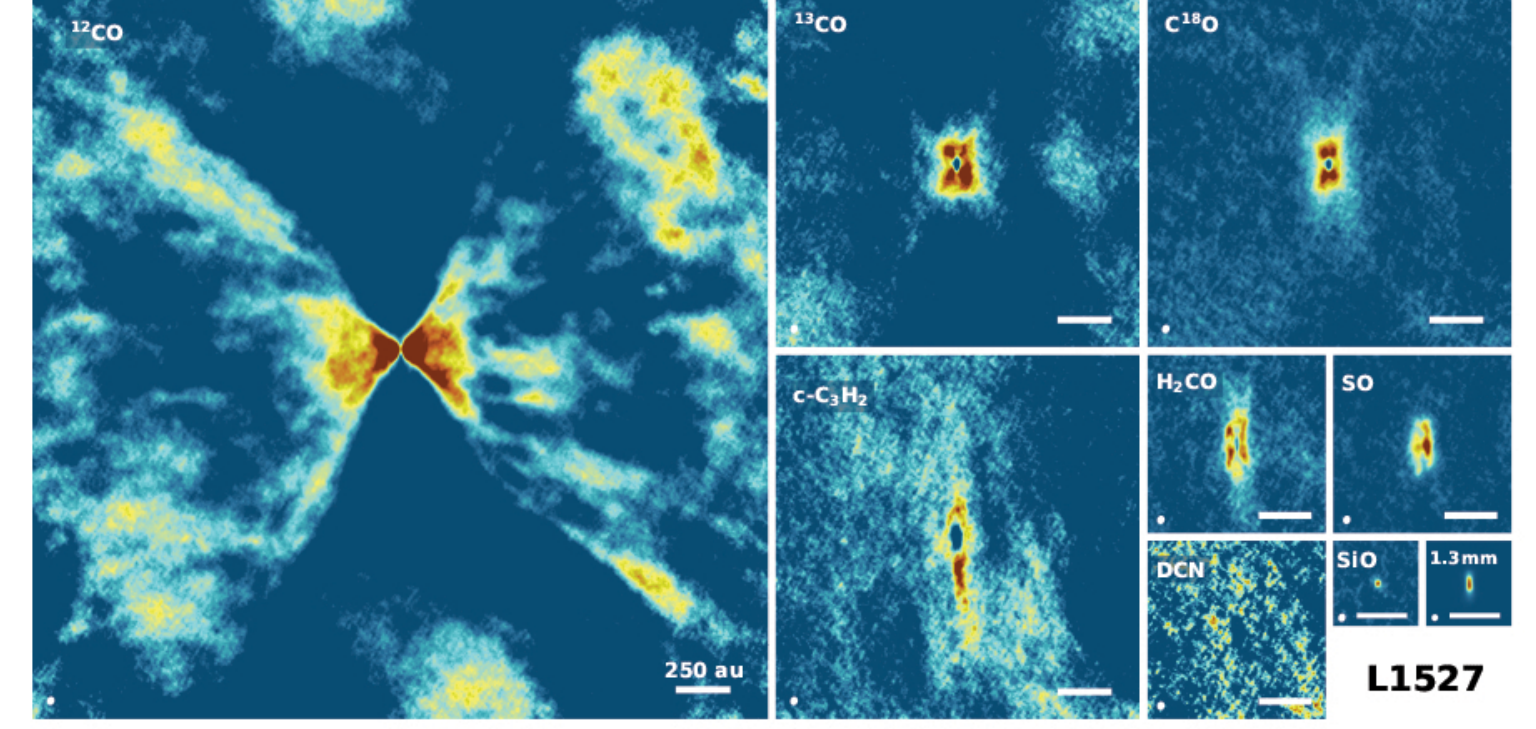}
\caption{Overview of 1.3 mm continuum and line observations (integrated intensity maps) toward L1527. For \HtCO the $3_{0,3}-2_{0,2}$ transition is shown, and for \cCtHt the blended $6_{0,6}-5_{1,5}$ and $6_{1,6}-5_{0,5}$ transitions are shown. The images are centered on the source position (R.A. = 04:39:53.9, Dec. = +26:03:09.4). The spatial scale is the same in each panel, with the white scale bar denoting 250 au. The size of the \twCO image is 24\asec $\times$ 24\asec. All color scales are linear, start at zero, and are saturated to highlight weaker, more extended emission. Intensity scales are shown in Fig.~\ref{fig:Molecularlines}. All images are made with a robust parameter of 2.0, and the beam size is depicted in the lower left corner of each panel. }
\label{fig:Prettygallery}
\end{figure*}


\section{Observations} \label{sec:Observations} 

L1527 has been observed as part of the ALMA Large Program eDisk (2019.1.00261.L; PI N. Ohashi) on 14 and 15 October 2022, sampling baselines between 91 and 11,469 m. Observations in a more compact configuration with the purpose of recovering larger spatial scales were carried out through a DDT program (2019.A.00034.S; PI J. Tobin) on 3 December 2021, 16 December 2021 and 3 July 2022. These observations used baselines between 15 and 2617 m. The correlator setup for both programs is centered around $\sim$225 GHz (1.3 mm) and includes 2 low spectral resolution windows at 976.56 kHz (1.34 \kms) resolution, 1 spectral window at 488.28 kHz (0.635 \kms) resolution, and 4 higher spectral resolution windows at 122.07 kHz (0.167 \kms) resolution. More details about the observations are provided by Ohashi et al. (subm.), and an overview of the molecular lines discussed in this paper is provided in Table~\ref{tab:molecularlines}.  

Standard calibration of the data was done using the ALMA Pipeline and a script developed for the eDisk Large Program, as described in Ohashi et al. (subm.) was used for subsequent data reduction and imaging. In short, the continuum emission peaks of all execution blocks were first aligned to a common phase center after which an amplitude rescaling was applied to the shifted visibilities. The amplitude calibration uncertainty is expected to be $\sim$5--10\%. Two rounds of continuum self-calibration were then performed on the aggregate continuum data, first on the short-baseline data only and then on the short- and long-baseline data combined. For L1527, phase and amplitude self-calibration were performed on the short baseline data, while only phase self-calibration was used for the combined data set. The final gain tables were also applied to the line data.

\begin{figure*}
\centering
\includegraphics[width=\linewidth,trim={0cm 13.7cm 0cm 1.0cm},clip]{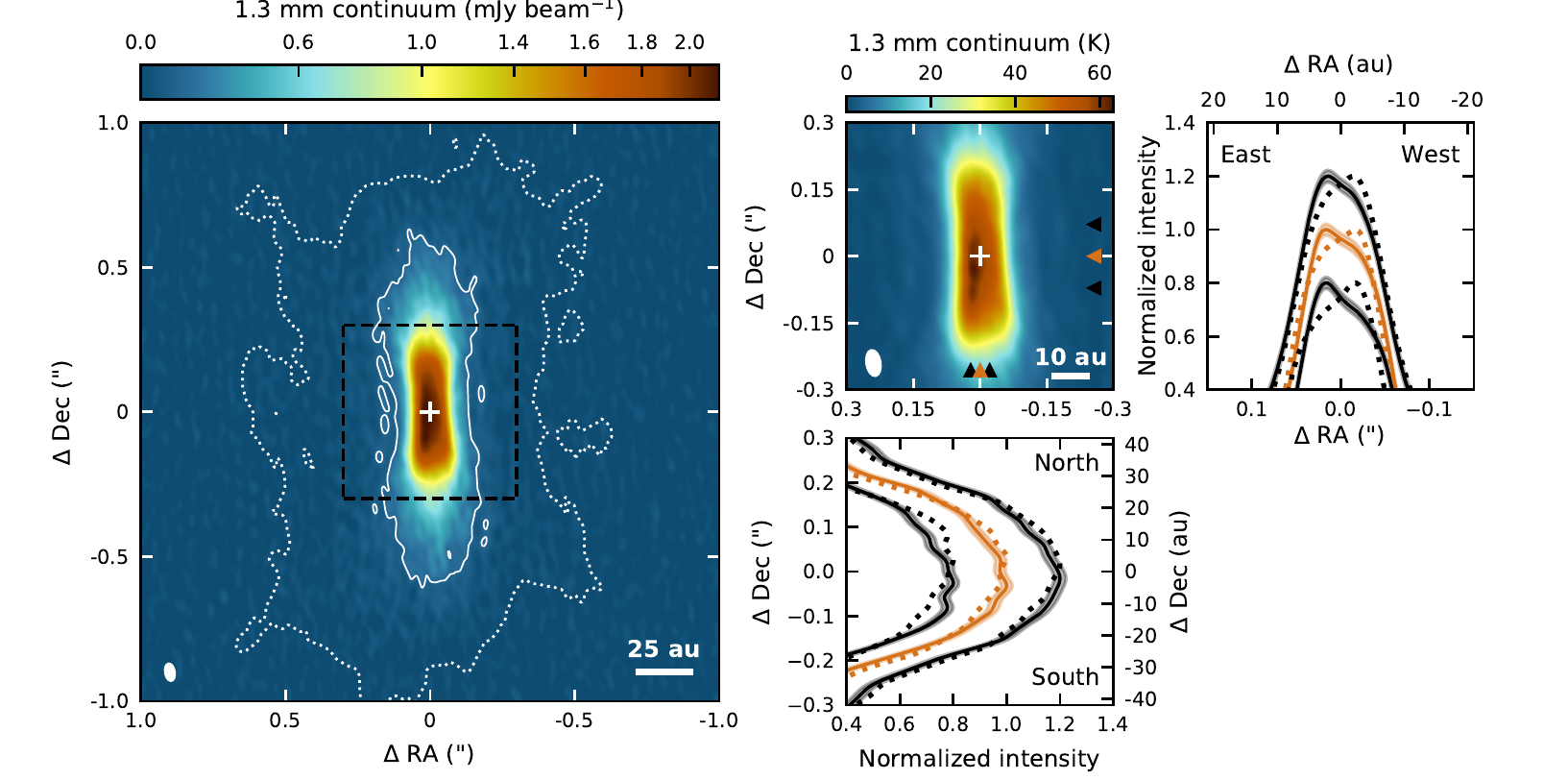}
\caption{ALMA 1.3 mm continuum image of L1527. The left panel shows the full extent of the continuum imaged with a robust parameter of $-$0.5 (color scale; beam size of 0\farcs056 $\times$ 0\farcs029, PA = 9.3$^{\circ}$), with the solid white contour marking the 5$\sigma$ level (0.15 mJy beam$^{-1}$). The dotted white contour marks the 5$\sigma$ level (0.11 mJy beam$^{-1}$) of the map imaged with a robust parameter of 2.0 (beam size of 0\farcs123 $\times$ 0\farcs111, PA = -13.7$^{\circ}$). The white cross marks the source position derived from a Gaussian fit in the image plane (R.A. = 04:39:53.9, Dec. = +26:03:09.4). The black dashed square shows the region depicted in the top middle panel, where the brightness temperature of the continuum is displayed. The top right panel shows the normalized intensity along the disk minor axis at source position (orange curve) and at 10 au to the north (top black curve) and south (bottom black curve). The bottom right panel shows the normalized intensity along the disk major axis at source position (orange curve) and at 3 au to the east (left black curve) and west (right black curve). Black and orange triangles in the continuum image (top middle panel) mark the locations of the intensity profiles. The black curves are shifted by 0.2 in intensity scale with respect to the orange curves for better visibility. The shaded region depicts the 3$\sigma$ level and the dotted lines are the mirror images of the solid lines to highlight the asymmetries. }
\label{fig:Continuum}
\end{figure*}

The standard eDisk image products were created with \texttt{tclean} using a range of robust parameters ($-$2.0, $-$1.0, $-$0.5, 0.0, 0.5, 1.0 and 2.0) for the continuum and robust = 0.5 for the line images. For L1527, line images were also created with a robust value of 2.0 to increase the signal-to-noise ratio, and those images are presented here. The resulting beam size for the line images is approximately 0\farcs17 $\times$ 0\farcs14 (PA = -20$^{\circ}$), and the noise levels for the different line cubes are listed in Table~\ref{tab:molecularlines}. Unless noted otherwise, we present the continuum image created with a robust value of $-$0.5 as a compromise between resolution and sensitivity. This image has a resolution of 0\farcs056 $\times$ 0\farcs029 (PA = 9.3$^{\circ}$), and an rms of 29 $\mu$Jy beam$^{-1}$. The full range of continuum images is presented in Fig.~\ref{fig:ContinuumOverviewRobust}. An overview of the continuum and molecular line observations toward L1527 is presented in Fig.~\ref{fig:Prettygallery}.

\begin{figure*}

\centering
\includegraphics[scale=1.05,trim={0.6cm 2.0cm 0.6cm 0.6cm},clip]{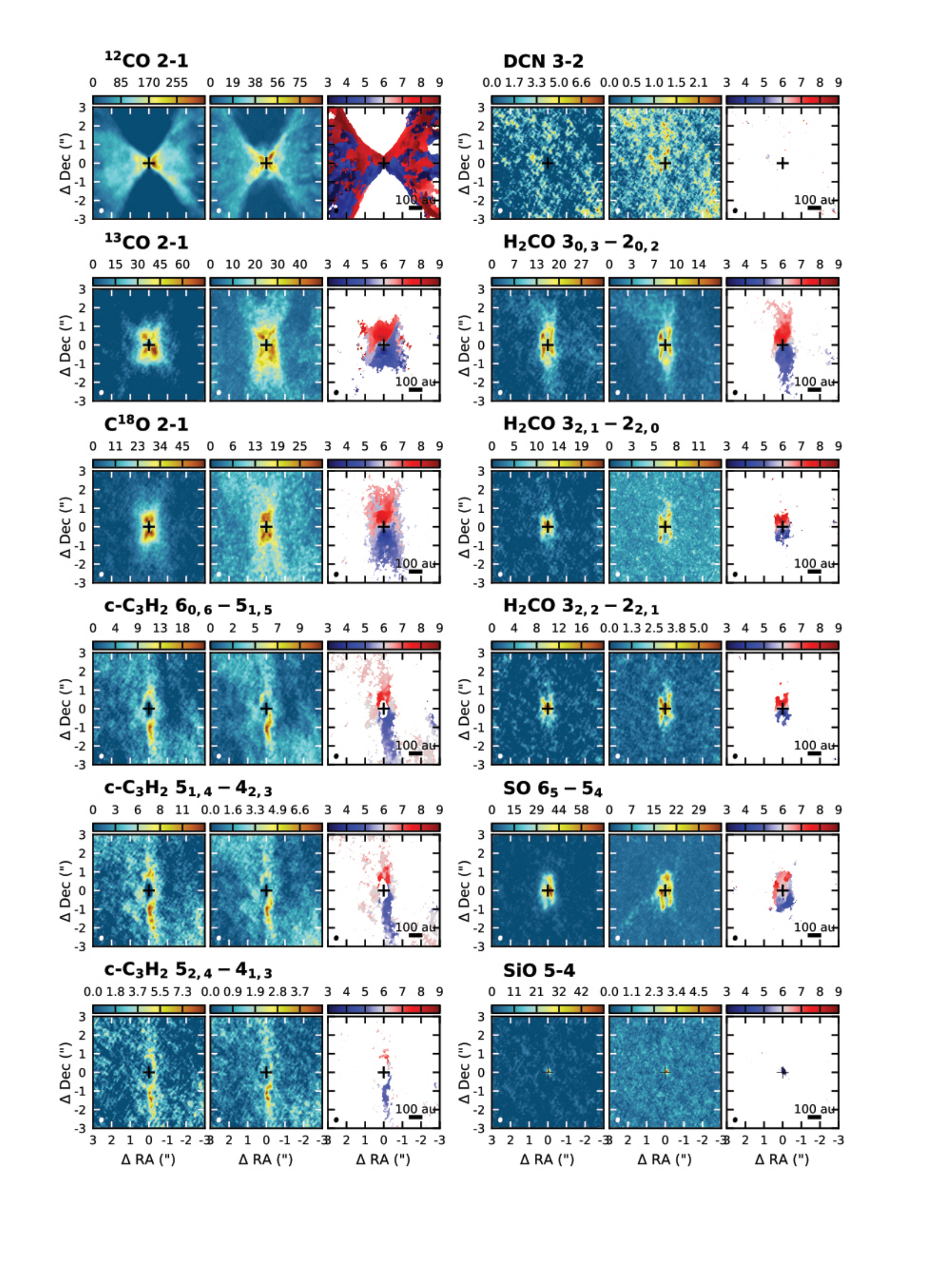}
\caption{Overview of molecular lines detected toward L1527, shown on scales of the disk and inner envelope. The first and fourth column present integrated intensity (moment zero) maps (in mJy beam$^{-1}$ \kms), the second and fifth column present peak intensity (moment eight) maps (in mJy beam$^{-1}$), and the third and sixth column present velocity maps made with \texttt{bettermoments} (in \kms, centered at the systemic velocity of 5.9 \kms). The velocity map is only showing pixels above the $> 3\sigma$ level of the moment zero map. The molecular line is indicated above the panels, and the beam is depicted in the lower left corner of each panel. }
\label{fig:Molecularlines}
\end{figure*}



\section{Results} \label{sec:Results}


\subsection{Continuum} \label{sec:Continuum}

The 1.3 mm continuum image of L1527 obtained with a robust parameter of $-$0.5 is shown in Fig.~\ref{fig:Continuum}, and a gallery of images made with different robust parameters is presented in Fig.~\ref{fig:ContinuumOverviewRobust}. The continuum image displays an edge-on disk with the major axis along the north--south direction, as previously observed \citep[e.g.,][]{Ohashi1997,Loinard2002,Tobin2008,Tobin2010,Tobin2012,Tobin2013,Sakai2014a,Aso2017,Nakatani2020,Ohashi2022,Sheehan2022}. The flared nature of the disk inferred by radiative transfer modeling of multi-wavelength continuum emission \citep{Tobin2013} is now clearly visible at this high resolution. The emission extends out to a radius of $\sim$0\farcs5 (70 au) along the major axis, and in the east--west direction to $\sim$0\farcs1 (14 au) above the midplane near the source position and up to $\sim$0\farcs2 (28 au) at a radial offset of $\sim$0\farcs3 (42 au). Fainter emission extending out to a radius of $\sim$1\asec (140 au) and to $\sim$0\farcs5 (70 au) above the midplane is visible in the image obtained with a robust parameter of 2.0. This is most likely due to more faint envelope emission being picked up by the higher sensitivity of the robust = 2.0 image, because the vertical extent of the bright central region is not much more extended (Fig.~\ref{fig:ContinuumOverviewRobust}). The brightness temperature for the majority of the disk is 40--60 K (for a robust parameter of $-$0.5), suggesting that the continuum may be optically thick. 

The continuum emission appears smooth with no sign of substructures. However, an asymmetry between the north and south side, with the south side being brighter than the north side, is visible for all robust parameters (see radial cuts for a robust value of $-$0.5 in Fig.~\ref{fig:Continuum}). At the highest resolutions (robust values $\le$ 0.0 or beam sizes of 0\farcs062$\times$0\farcs038 and smaller), an asymmetry between the east and west side becomes also clear, with the east side being brighter than the west side. The north--south asymmetry is then also more pronounced in the east. The east--west asymmetry is stronger in the south but visible along the entire major axis. The difference in maximum brightness temperature between the southeast and the northeast or southwest is about 10 K. 

Because of the flared nature of the disk, a simple 2D Gaussian fit does not represent the emission morphology and total flux density. Therefore, we sum over all pixels with values $> 3\sigma$, which yields a flux density of 192.57 $\pm$ 0.05 mJy (compared to 139.04 $\pm$ 0.73 mJy from the Gaussian fit). This is only 9\% higher than the 1.3 mm flux reported by \citet{Aso2017} obtained within a 4\asec $\times$ 4\asec box from 0\farcs47 $\times$ 0\farcs37 resolution observations. Using a similar aperture, we obtain a slightly lower flux density of 188.78 mJy, which is within 7\% of the previously reported value.

For isothermal and optically thin emission, the continuum flux density, $S_\nu$, can be converted into a dust mass using 
\begin{equation}
    M_{\rm{dust}} = \frac{D^2S_\nu}{\kappa_\nu B_\nu(T_{\rm{dust}})},  
\end{equation}
where $D$ is the distance (140 pc), $\kappa_\nu$ the dust opacity at the observed frequency, and $B_\nu(T_{\rm{dust}})$ the Planck function at the observed frequency for a dust temperature of $T_{\rm{dust}}$. We adopt a dust opacity of 2.3 cm$^2$ g$^{-1}$ at the observed frequency of 225 GHz \citep{Beckwith1990}, and an average temperature of \mbox{$T = 43 \,(L/L_\Sun)^{0.25}$ K = 46 K} (with $L$ = 1.3 $L_\Sun$; Ohashi et al. subm.), based on a suite of radiative transfer models by \citet{Tobin2020}. This results in a dust mass of $\sim$41 $M_\Earth$. Adopting a temperature of 20 K, as typically done for Class II disks \citep[e.g.,][]{Ansdell2016}, results in a dust mass of $\sim$112 $M_\Earth$. For a gas-to-dust ratio of 100, the total disk mass is then $\sim$0.01--0.03 $M_\Sun$, consistent with previous estimates at different wavelengths given different choices for temperature and dust opacity \citep[e.g.][]{Tobin2013,Aso2017,Nakatani2020,Sheehan2022}. The here derived mass is a lower limit because the continuum emission is likely optically thick.

\begin{figure*}[t]
\centering
\includegraphics[width=\linewidth,trim={1cm 11.3cm 0.5cm 0.6cm},clip]{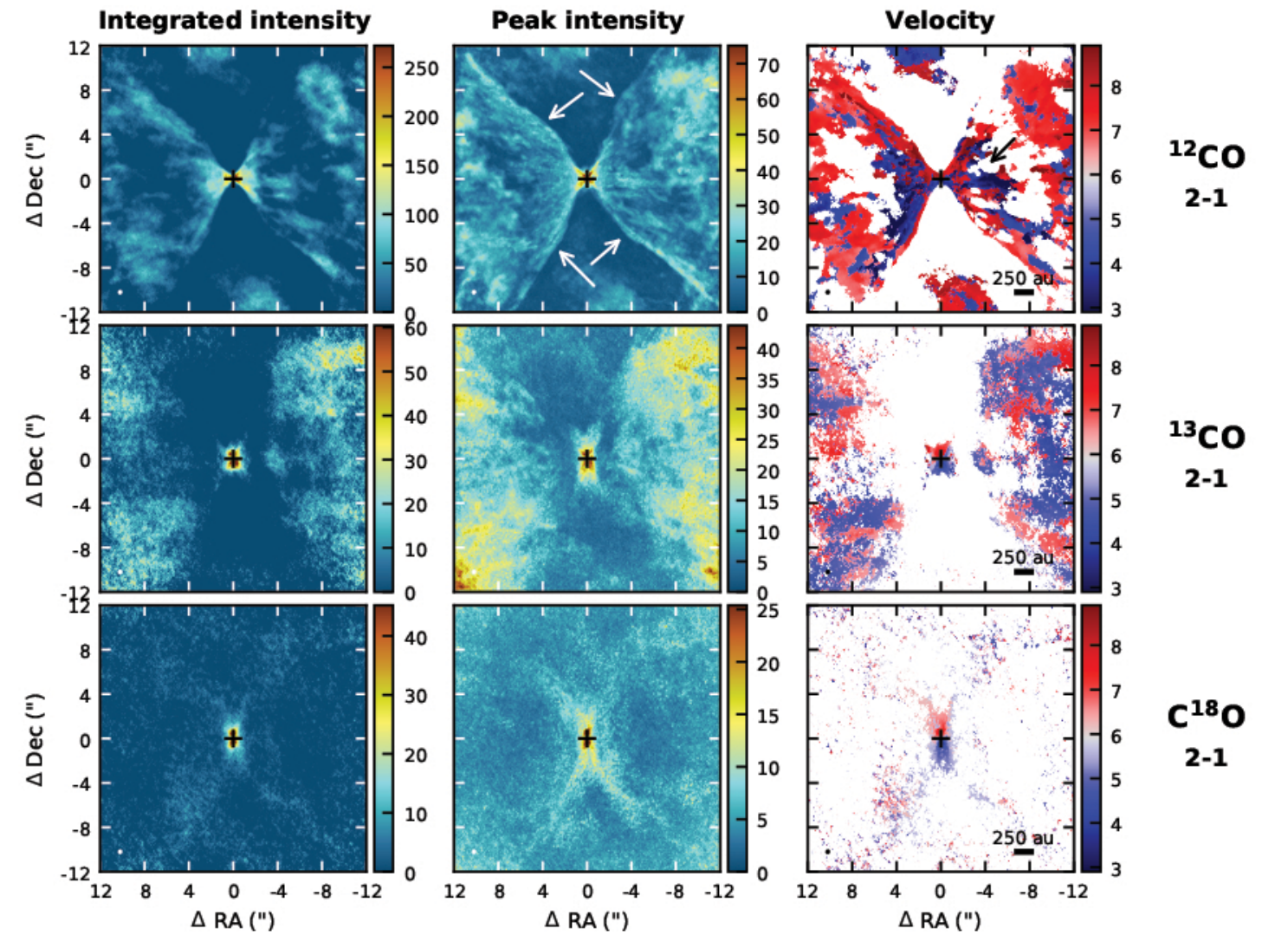}
\caption{Overview of CO isotopologue emission at scales larger than depicted in Fig.~\ref{fig:Molecularlines}. The first column presents integrated intensity (moment zero) maps (in mJy beam$^{-1}$ \kms), the middle column presents peak intensity (moment eight) maps (in mJy beam$^{-1}$), and the right column presents velocity maps made with \texttt{bettermoments} (in \kms, centered at the systemic velocity of 5.9 \kms). The color scale of the moment zero and moment eight maps are capped to highlight emission at large scales. The velocity map is only showing pixels above the $> 3\sigma$ level of the moment zero map. The beam is shown in the bottom left corner of each panel. The white arrows in the \twCO peak intensity map (top middle panel) highlight the kinks discussed in the main text (Sect.~\ref{sec:CO}), and the black arrow in the \twCO velocity map marks the potential jet.}
\label{fig:CO_maps}
\end{figure*}


\subsection{Molecular lines} \label{sec:Lines}

An overview of all molecular lines detected towards L1527 is listed in Table~\ref{tab:molecularlines} and presented in Fig.~\ref{fig:Molecularlines}. In addition to moment zero maps (integrated intensity), we show moment eight maps (peak intensity) and velocity maps (similar to moment nine maps) created with the quadratic method of the \texttt{bettermoments} package\footnote{https://github.com/richteague/bettermoments} \citep{Teague2018}. It is evident from Fig.~\ref{fig:Molecularlines} that each molecular species and for H$_2$CO, each transition, exhibits a different spatial and velocity distribution. Each species and its spatial origin (e.g., outflow, envelope, disk) is therefore discussed individually below (Sect.~\ref{sec:CO}--\ref{sec:SO}), before the full molecular structure and the underlying physical and/or chemical structure are discussed in Sects.~\ref{sec:Temperature} and \ref{sec:MolecularStructure}.


\subsubsection{$^{12}$CO, $^{13}$CO, C$^{18}$O} \label{sec:CO}

In addition to the images on disk-scales shown in Fig.~\ref{fig:Molecularlines}, moment zero, moment eight and velocity maps on larger scales are presented in Fig.~\ref{fig:CO_maps} for the CO isotopologues. The large-scale \twCO emission displays an hour glass morphology, with emission originating along and inside the outflow cavity walls. The cavity walls deviate from a parabolic shape and display a kink (highlighted with white arrows in Fig.~\ref{fig:CO_maps}). This kink occurs at larger offsets from the source for the northwestern and southeastern cavity walls. The most prominent feature inside the cavity walls is visible $\sim$3--6\asec (420--840 au) west of the source position at blueshifted velocities (highlighted with an arrow in the velocity map in Fig.~\ref{fig:CO_maps} and in the channel maps in Fig.~\ref{fig:12CO_channels}).

\twCO emission is detected over a velocity range of $-$11.3 -- $-$1.13 and 0.77 -- 9.04 \kms (with respect to the systemic velocity of 5.9 \kms \citealt{Caselli2002a,Tobin2011}), while most of the emission is resolved out at velocities close to the systemic velocity. There is no clear velocity gradient visible in the outflow direction (east--west), but the southern outflow cavity walls are more pronounced at blueshifted velocities, while the northern cavity walls are stronger at redshifted velocities. This velocity pattern is more clearly visible at smaller scales (Fig.~\ref{fig:Molecularlines}) and is similar to the rotation direction of the disk and inner envelope. 

The \thCO and \CeO emission are dominated by the disk and inner envelope (Fig.~\ref{fig:Molecularlines}), but they also have a contribution from material inside and along the cavity walls. Emission inside the cavity walls is clearly visible in the \thCO images in Fig.~\ref{fig:CO_maps} (at offsets larger than $\sim$4\asec in the east--west direction), while for \CeO it is only visible in the individual velocity channels (see Fig.~\ref{fig:C18O_channels}). Large-scale emission is seen out to velocity offsets of $\sim |7.5|$ \kms in \twCO emission, while this is only visible out to $\sim |2.0|$ \kms in \thCO and \CeO, maybe due to the lower sensitivity at the higher velocity resolution. For both \thCO and \CeO, there is a narrow arc of emission in both outflow cavities that moves outward with increasing velocity offsets (Figs.~\ref{fig:13CO_channels} and \ref{fig:C18O_channels}). A similar moving ``front'' of emission is also visible in a range of $^{12}$CO channels (Fig.~\ref{fig:12CO_channels}), but at smaller spatial scales ($\sim$2--3\asec off source, compared to $\sim$2--$>$16\asec for \thCO and \CeO) and higher velocity offsets ($\sim |2.5-7|$ \kms, compared to $\lesssim |2.0|$ \kms for \thCO and \CeO). This differences between \twCO and \thCO/\CeO is likely because most \twCO emission is resolved out at the velocities where the \thCO/\CeO moving ``front'' is detected and the sensitivity is not high enough to detect \thCO/\CeO emission at velocities as high as for \twCO.  

\begin{figure*}
\centering
\subfloat{\includegraphics[width=\linewidth,trim={0.5cm 19.5cm 1.7cm 0.4cm},clip]{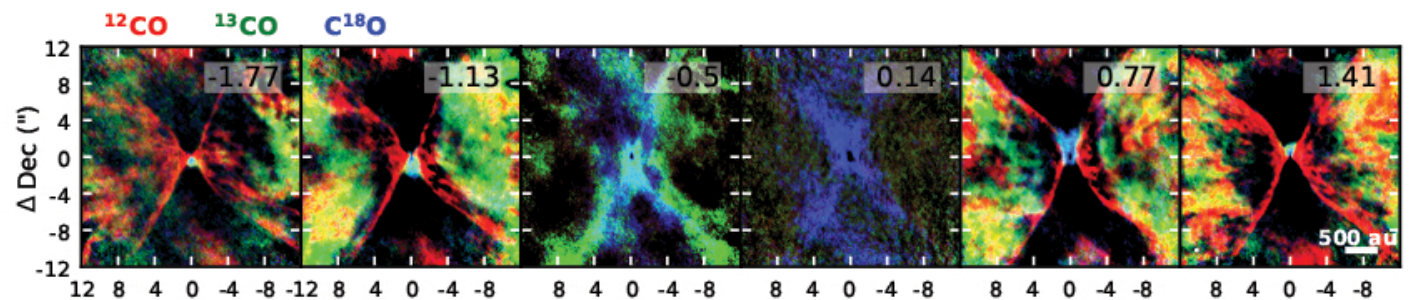}}\\
\subfloat{\includegraphics[width=\linewidth,trim={0.5cm 19cm 1.7cm 0.7cm},clip]{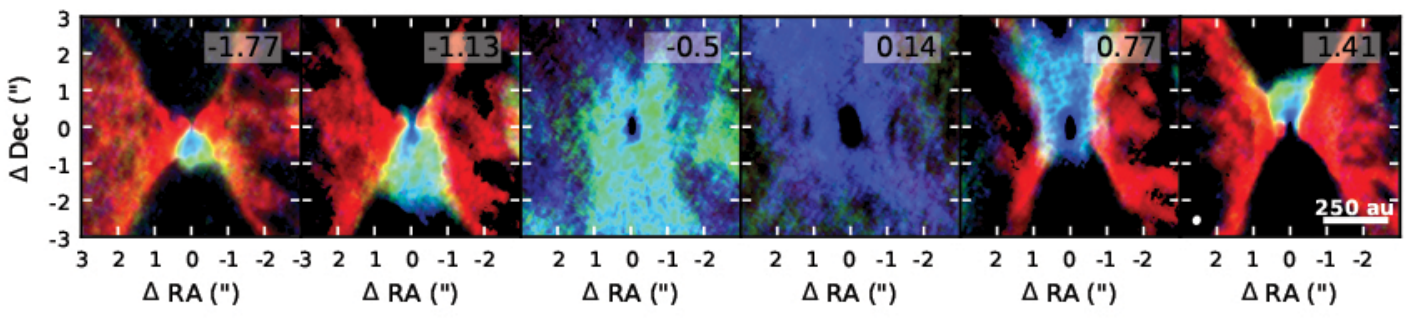}}
\caption{Selected velocity channel maps showing $^{12}$CO (red), $^{13}$CO (green) and C$^{18}$O (blue) emission overlaid. The bottom row is zoomed in on the central part of the channels shown in the top row. The velocity with respect to the systemic velocity is listed in the top right corner of each panel. The beam (nearly identical for all three isotopologues) is shown in the bottom left corner of the right most panels. The same channels are shown for each individual molecule in Fig.~\ref{fig:CO_channels}.}
\label{fig:CO_channels_overlay}
\end{figure*}

The large-scale \CeO moment eight map (Fig.~\ref{fig:CO_maps}) displays X-shaped emission ($\lesssim$8\asec), while very faint and narrow X-shaped emission is visible in some \thCO velocity channels (Fig.~\ref{fig:CO_channels}) and very weakly in the moment zero map on smaller spatial scales ($\lesssim$4\asec). This structure was previously observed for \thCO $J=1-0$ emission \citep{Ohashi1997}. One of the reasons for the difference between the \thCO and \CeO moment maps is that while \thCO (and \twCO) emission is resolved out near the systemic velocity, \CeO is detected in all low-velocity channels. Therefore, to better compare the spatial origin of the emission from the different CO isotopologues, \thCO and \CeO images are created at the same velocity resolution as the \twCO image, and velocity channels with emission from all three isotopologues overlaid are presented in Fig.~\ref{fig:CO_channels_overlay}. Channel maps for each individual isotopologue are shown in Fig.~\ref{fig:CO_channels}. The overlay in Fig.~\ref{fig:CO_channels_overlay} reveals a layered structure, with \twCO tracing the surface layer of the cavity wall and \thCO and \CeO tracing deeper and deeper into the envelope. The faint and narrow X-shaped emission visible in \thCO channels coincides with \twCO emission from the cavity wall (visible in yellow), while the broad X-shape in the \CeO moment maps is caused by emission at velocities close to the systemic velocity. A similar structure is visible for \thCO at $-$0.5 \kms, and this velocity channel clearly shows that the \thCO emission originates from layers closer to the outflow cavity. The emission from the inner envelope and disk, as traced by \thCO and \CeO ($-$1.77, $-$1.13, 0.77, 1.41 \kms in Fig.~\ref{fig:CO_channels_overlay}), is present in between the \twCO emission features. Here too, the \thCO emission is peaking in higher layers compared to \CeO. The \CeO channel maps at the original velocity resolution (Fig.~\ref{fig:C18O_channels}) clearly show that the emission has a contribution from outflowing material as well as from the surface layer of the envelope or cavity wall. At velocities close to the systemic velocity ($\pm$ $\sim$0.5 \kms), these components start to overlap. 

\begin{figure*}
\centering
\includegraphics[width=\linewidth,trim={0cm 9.2cm 0cm 6.8cm},clip]{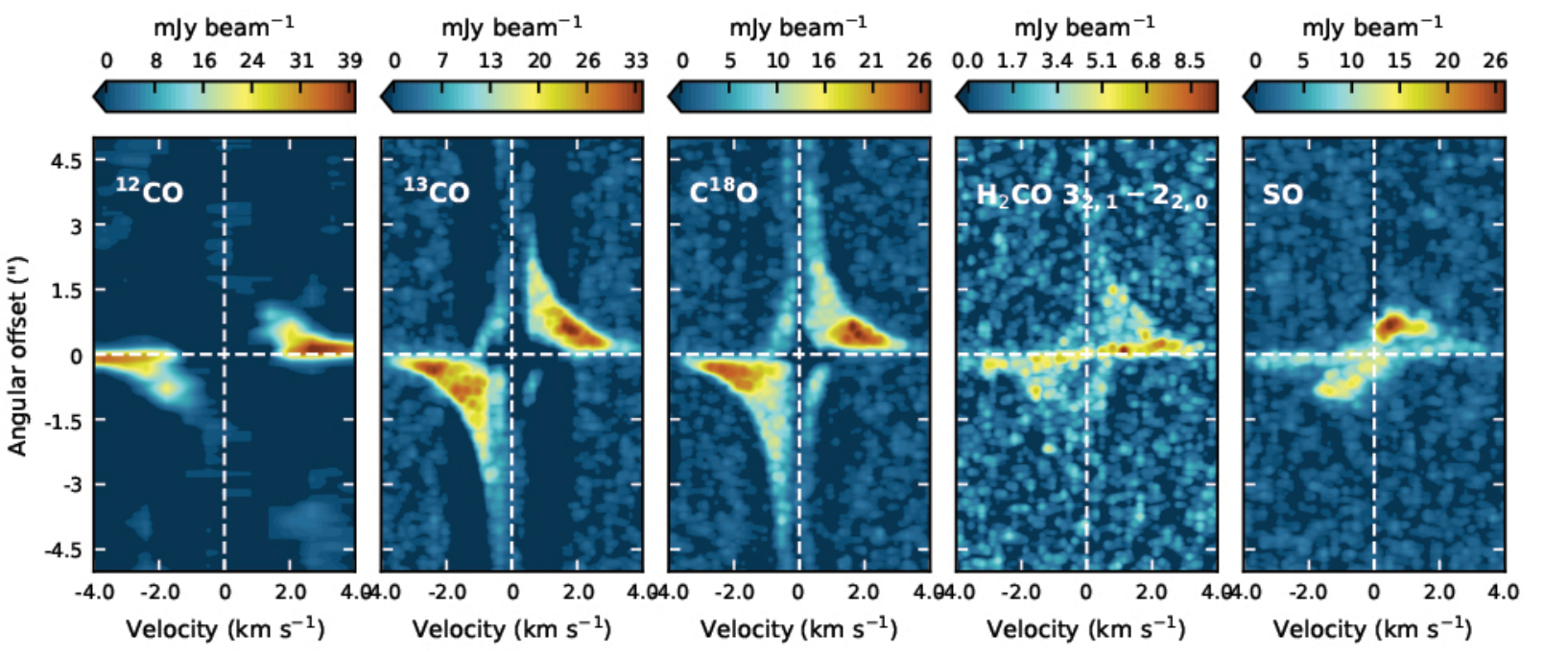}
\caption{Position-velocity diagrams extracted along the major axis (averaged over the size of the beam) for \twCO (first panel) and the molecular lines observed at 0.167 \kms resolution: \thCO (second panel), \CeO (third panel), \HtCO $3_{2,1}-2_{2,0}$ (fourth panel), and SO (fifth panel). For \twCO, a cropped velocity range is shown for better visibility of the structure of the other molecular lines. The horizontal and vertical dotted white lines mark the source position and systemic velocity (which is shifted to 0 \kms), respectively. }
\label{fig:pvdiagrams}
\end{figure*}

While both \thCO and \CeO trace emission from the disk and inner envelope, the contributions from the different components are not exactly the same for both isotopologues. The difference in origin between \thCO and \CeO emission becomes more clear from position-velocity (pv) diagrams, as presented in Fig.~\ref{fig:pvdiagrams}. Emission at angular offsets $\gtrsim$2\asec as well as emission in the non-Keplerian quadrants (top left and bottom right) is relatively stronger for \CeO than for \thCO. Together with the fact that \thCO emission is resolved out in the central channel while \CeO emission is not, this suggests that \CeO traces emission from the infalling envelope out to larger scales than \thCO emission. This is probably because \thCO becomes optically thick faster due to its higher abundance and hence gets resolved out more heavily near line center. Close to the central protostar, the \CeO emission is concentrated on scales $\lesssim$ 1\asec and velocities $\gtrsim$ 1 \kms, while \thCO emission extends out to slightly larger scales and slightly lower velocities. This suggests that the contribution from the innermost envelope is stronger for \thCO than it is for \CeO, likely due to the higher \thCO abundance. A similar conclusion can be drawn from the velocity maps (Fig.~\ref{fig:Molecularlines}), where \thCO displays a stronger contribution from blueshifted envelope emission in the north and redshifted envelope emission in the south, that is, opposite to the disk velocity structure. 

The intensity ratio between \thCO and \CeO in channels with disk and inner envelope emission ($|\Delta v| \gtrsim$ 1 \kms) is $\sim$1--3 instead of the canonical ratio of $\sim$7 \citep{Wilson1994}. This suggests that \thCO is generally optically thick and that \CeO is probably optically thick in the midplane region where the line ratio is lowest ($\sim$1), consistent with earlier observations \citep{vantHoff2018b}. The central negative ``gap'' visible in the \thCO and \CeO moment zero map (and low-velocity channels, e.g., Fig.~\ref{fig:CO_channels_overlay} and most clearly visible in Fig.~\ref{fig:CO_channels}) is therefore likely the result of continuum oversubtraction from optically thick line emission, potentially in combination with absorption by cold outer envelope material that is being resolved out.

\subsubsection{SiO}

\begin{figure*}
\centering
\includegraphics[width=\linewidth,trim={0cm 16.5cm 0cm 1.0cm},clip]{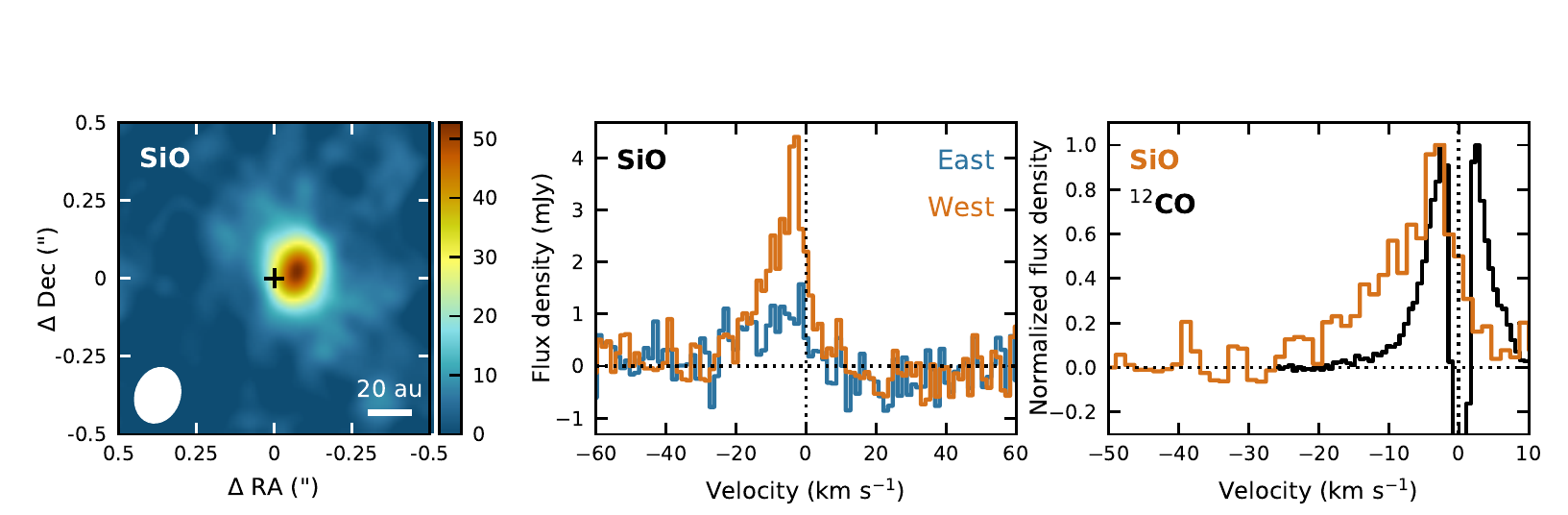}
\caption{Integrated intensity (moment zero) map of SiO in the inner 0\farcs5 (left panel) and spectra extracted in a 0\farcs2 aperture (middle panel) centered on the SiO peak 0\farcs08 west of the source position (orange) and at the same angular offset to the east (blue). In the right panel the normalized spectrum toward the SiO peak (orange) is compared to the normalized \twCO spectrum extracted in the same aperture (black). The color scale of the integrated intensity map is in mJy beam$^{-1}$ km s$^{-1}$. The beam is depicted in the lower left corner and the cross marks the source position.  }
\label{fig:SiO_12CO}
\end{figure*}

Very compact, mostly unresolved SiO emission is detected just west of the source position, peaking at an offset of $-$0\farcs08 ($\sim$11 au; Fig~\ref{fig:SiO_12CO}). The emission is predominantly blueshifted and is detected at velocities ranging between $-$13.45 and 1.30 \kms with respect to the systemic velocity, and peaks at $-$2.72 \kms. This component is not seen in any of the other molecular lines. \twCO emission typically peaks towards the northwest or southwest, rather than directly west, and emission close to source is only detected out to $-$10.18 \kms in individual velocity channels (see also Fig.~\ref{fig:SiO_12CO}, right panel). SiO is a shock tracer and is typically observed in protostellar jets. A jet origin of the SiO emission observed toward L1527 is consistent with the high velocities of the emission.


\subsubsection{c-C$_3$H$_2$}

All four \cCtHt transitions (two are blended, see Table~\ref{tab:molecularlines}) display strong emission features along the north--south direction out to offsets of $\sim$2\asec (280 au; Fig.~\ref{fig:Molecularlines}), which is more extended than the bright components seen in \thCO and \CeO ($\sim$1\asec). The blueshifted emission in the south is stronger than the redshifted emission in the north. In contrast, the total extent in the east--west direction is only $\sim$0\farcs3, compared to $\sim$1\farcs0 and $\sim$2\farcs0 for \CeO and \thCO, respectively.  In addition, a weak large scale emission component is visible (most clearly seen in Fig.~\ref{fig:Prettygallery}).  

At the velocity resolution of 1.34 \kms, the \cCtHt transitions are detected in only three channels, spanning a velocity range of $-$1.16 -- 1.53 \kms, except for the blended $6_{0,6}-5_{1,5}$ and $6_{1,6}-5_{0,5}$ transitions that display weak (3--4$\sigma$) emission at $-$2.5 \kms (Fig.~\ref{fig:c-C3H2_channels}). The faint extended component is only visible in the central channel. The combination of emission morphology and narrow velocity range indicates that \cCtHt is tracing envelope emission. A central absorption feature is only present in the central velocity channel and is likely due to absorption of the warm continuum emission by cold envelope material.

\subsubsection{DCN}

Very weak DCN emission is tentatively detected at the 3--4$\sigma$ level in the central velocity channel (Fig.~\ref{fig:DCN_channels}). The emission is extended surrounding a central region with negative emission and bears the most resemblance to the faint large scale emission component seen in \cCtHt (as shown in Fig.~\ref{fig:Prettygallery}). There is a hint of a narrow emission feature north of the source extending in the north--south direction in the first redshifted channel (1.53 \kms), but such a feature is absent in the south in the first blueshifted channel (-1.16 \kms). This is reflected in the peak intensity map shown in Fig.~\ref{fig:Molecularlines}. DCN is thus at least present in the envelope and the central absorption feature is likely due to absorption of the warm continuum.


\subsubsection{H$_2$CO}

Three \HtCO transitions are observed, with upper level energies of 21 K ($3_{0,3}-2_{0,2}$) and 68 K ($3_{2,1}-2_{2,0}$ and $3_{2,2}-2_{2,1}$). The low-energy transition displays stronger and more extended emission than the higher energy transitions. All transitions display X-shaped emission, extending about 0\farcs8 toward the north and south, and 0\farcs4 toward the east and west (Fig.~\ref{fig:Molecularlines}; most clearly visible in the peak intensity maps). The low-energy transition ($3_{0,3}-2_{0,2}$) shows an emission peak along the north--south axis at $\sim$1\asec offsets. This feature is smaller and less evident for the higher energy transitions. Only the $3_{0,3}-2_{0,2}$ transition displays weak extended emission in the north--south direction as well as inside the outflow cavity walls ($\gtrsim$8\asec). 

\begin{figure*}
\centering
\includegraphics[width=\linewidth,trim={0.5cm 13.3cm 0.5cm 0.4cm},clip]{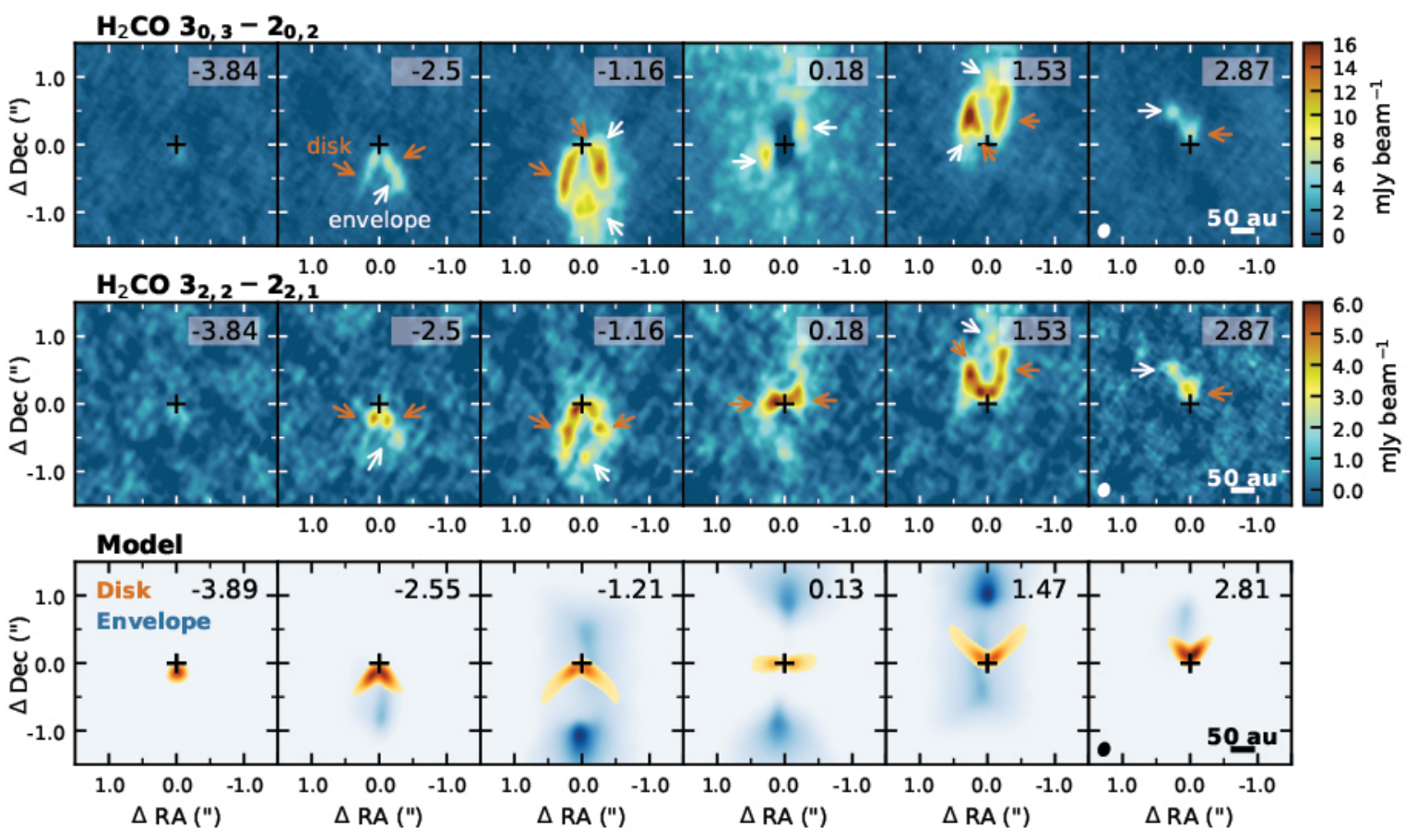}
\caption{Velocity channel maps for the \HtCO $3_{0,3}-2_{0,2}$ (top row) and $3_{2,2}-2_{2,1}$ (middle row) transitions, and for a model of a 125 au (radius) Keplerian rotating disk embedded in a rotating infalling envelope with molecular line emission originating in the disk surface layers and the inner envelope (bottom row; \citealt{Tobin2013,vantHoff2018b,vantHoff2020,vantHoff2022}). In the model panels, the contribution from the disk is shown in an orange color gradient and the envelope contribution is shown in a blue color gradient. For visualization purposes, the disk emission is depicted on top of the envelope emission. The model is meant as a qualitative comparison to determine the origin of the \HtCO emission. Observed emission features most likely originating in the disk surface are marked by orange arrows, and envelope features by white arrows. The black cross marks the source position. The velocity with respect to the systemic velocity is listed in the top right corner of each panel. The beam is shown in the bottom left corner of the right most panels.}
\label{fig:H2CO_channels}
\end{figure*}

The difference in emission morphology between the \HtCO transitions, as well as the origin of the emission can be more clearly seen in the individual velocity channels. In Fig.~\ref{fig:H2CO_channels} we compare the $3_{0,3}-2_{0,2}$ and $3_{2,2}-2_{2,1}$ transitions because they are observed at the same low velocity resolution of 1.34 \kms. In order to provide a qualitative description of the origin of the \HtCO emission we also present velocity channel maps of a model with a Keplerian disk (125 au radius) embedded in a rotating and infalling  envelope (CMU; \citealt{Ulrich1976,Cassen1981})\footnote{The CMU model describes the collapse of an isothermal, spherically symmetric, and uniformly rotating cloud. The assumption of free fall with each particle conserving angular momentum results in particles moving along parabolic paths. Details about the implementation can be found in Appendix C of \citet{vantHoff2018b}.}. The temperature and density structure are based on radiative transfer modeling of multi-wavelength continuum observations of L1527 by \citet{Tobin2013} and this model has been used by \citet{vantHoff2018b,vantHoff2020,vantHoff2022} to study molecular line observations toward L1527. The model images are created with the radiative transfer code LIME \citep{Brinch2010}. Our goal here is not to reproduce the observed emission exactly, but to illustrate the emission features expected for emission originating in different parts of the protostellar system. The model presented in Fig.~\ref{fig:H2CO_channels} has emission originating in the inner 250 au of the envelope and in the surface layers of the disk.  

The low-energy $3_{0,3}-2_{0,2}$ transition is detected over a velocity range of $-$3.84 -- 2.87 \kms, while the high-energy transition is only marginally detected at $-$3.84 \kms (Fig.~\ref{fig:H2CO_channels}). The $3_{2,1}-2_{2,0}$ transition is observed at a higher velocity resolution (0.167 \kms) and emission is detected over a velocity range of $-$3.05 -- 3.14 \kms (Fig.~\ref{fig:H2CO_channels_full}), suggesting that the asymmetry in velocity for the other transitions is likely due to the low spectral resolution. The V-shaped emission pattern responsible for the X-shape in the integrated intensity map (Fig.~\ref{fig:Molecularlines}) is visible in all channels, except the central channel, and is characteristic of emission originating in the surface layers of the disk. In contrast, the emission features $\sim$1\asec north and south of the source position seen most strongly for $3_{0,3}-2_{0,2}$ are only present in the $-$1.16 and 1.53 \kms channels and not at higher velocities, and likely originate in the inner envelope. Both transitions differ in the central velocity channel; where the $3_{2,2}-2_{2,1}$ emission extends to the east and west in a bow-tie pattern, the $3_{0,3}-2_{0,2}$ transition shows absorption surrounded by compact emission features to the southeast and northwest. The bow-tie pattern of $3_{2,2}-2_{2,1}$ is indicative of disk emission, while the asymmetric features, as well as the more extended features at low velocities in the north and south of $3_{0,3}-2_{0,2}$, are consistent with envelope emission. The $3_{2,2}-2_{2,1}$ bow-tie extends as far in the east and west direction as the $3_{0,3}-2_{0,2}$ emission features, so the absence of the bow-tie for the $3_{0,3}-2_{0,2}$ transition is not due to the central absorption. The warm $3_{2,2}-2_{2,1}$ transition thus originates predominantly in the disk surface layers, while the colder $3_{0,3}-2_{0,2}$ transition also has a strong contribution from the envelope. Absorption of continuum emission by the outer envelope is then the likely origin of the absorption in the central velocity channel of the $3_{0,3}-2_{0,2}$ transition.

\begin{figure*}
\centering
\includegraphics[width=\linewidth,trim={0.5cm 14.5cm 0.5cm 0.4cm},clip]{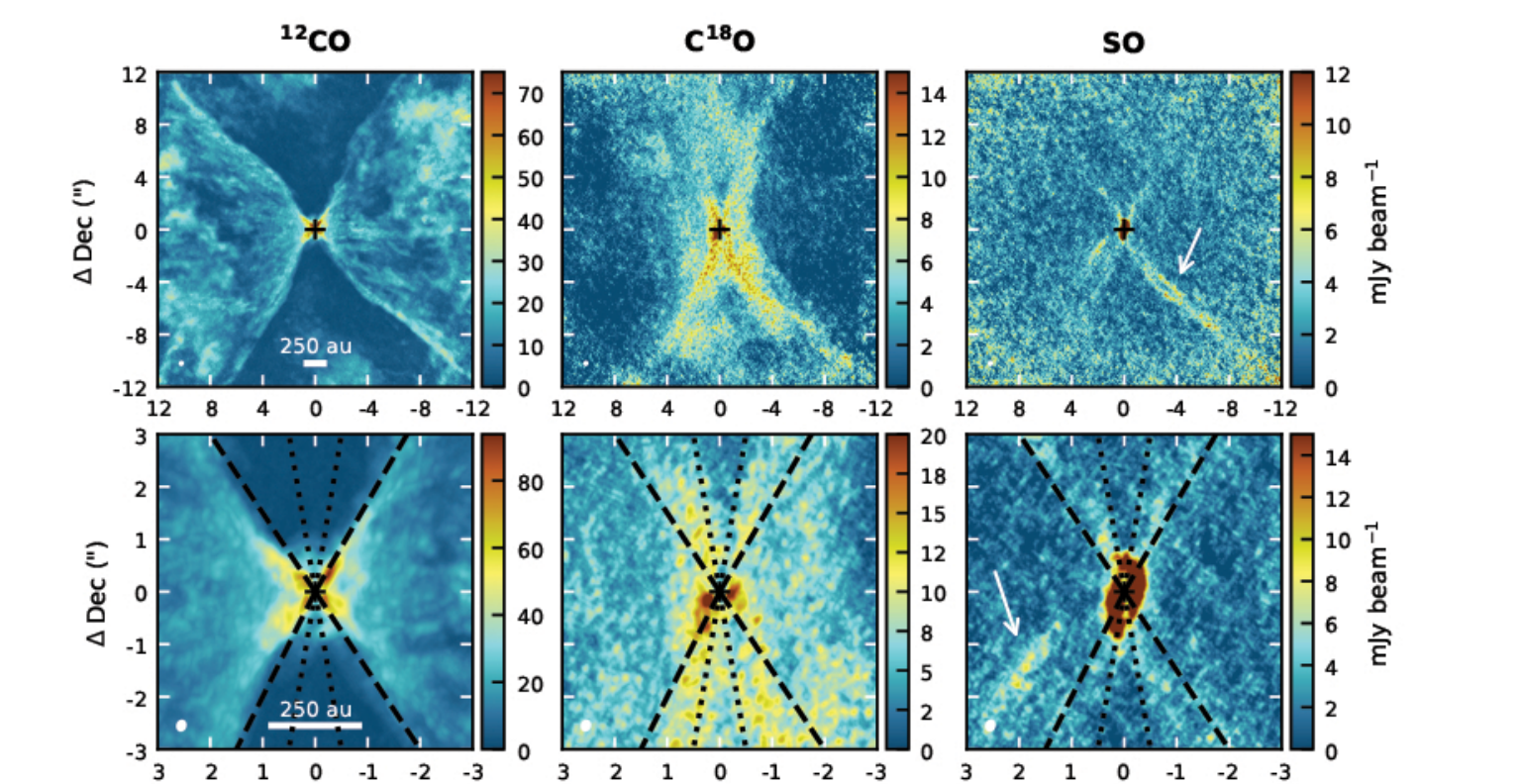}
\caption{Moment eight (peak intensity) maps for \twCO (left panels), \CeO (middle panels) and SO (right panels) on a scale of 24\asec (top panels) and 6\asec (bottom panels). For \CeO and SO, only the five central velocity channels ($-$0.37 -- 0.29 \kms) are used to highlight the large-scale emission features, while for \twCO the full velocity range ($-$11.3 -- 9.04 \kms) has been used. The dashed and dotted lines are the same in all three bottom panels and are meant to help guide the eye with respect to the location of the emission. The white arrows point to features described in the main text. The beam is shown in the bottom left corner of each panel. }
\label{fig:12CO_C18O_SO}
\end{figure*}

\begin{figure*}
\centering
\includegraphics[trim={0.0cm 1.4cm 0.5cm 0.0cm},clip]{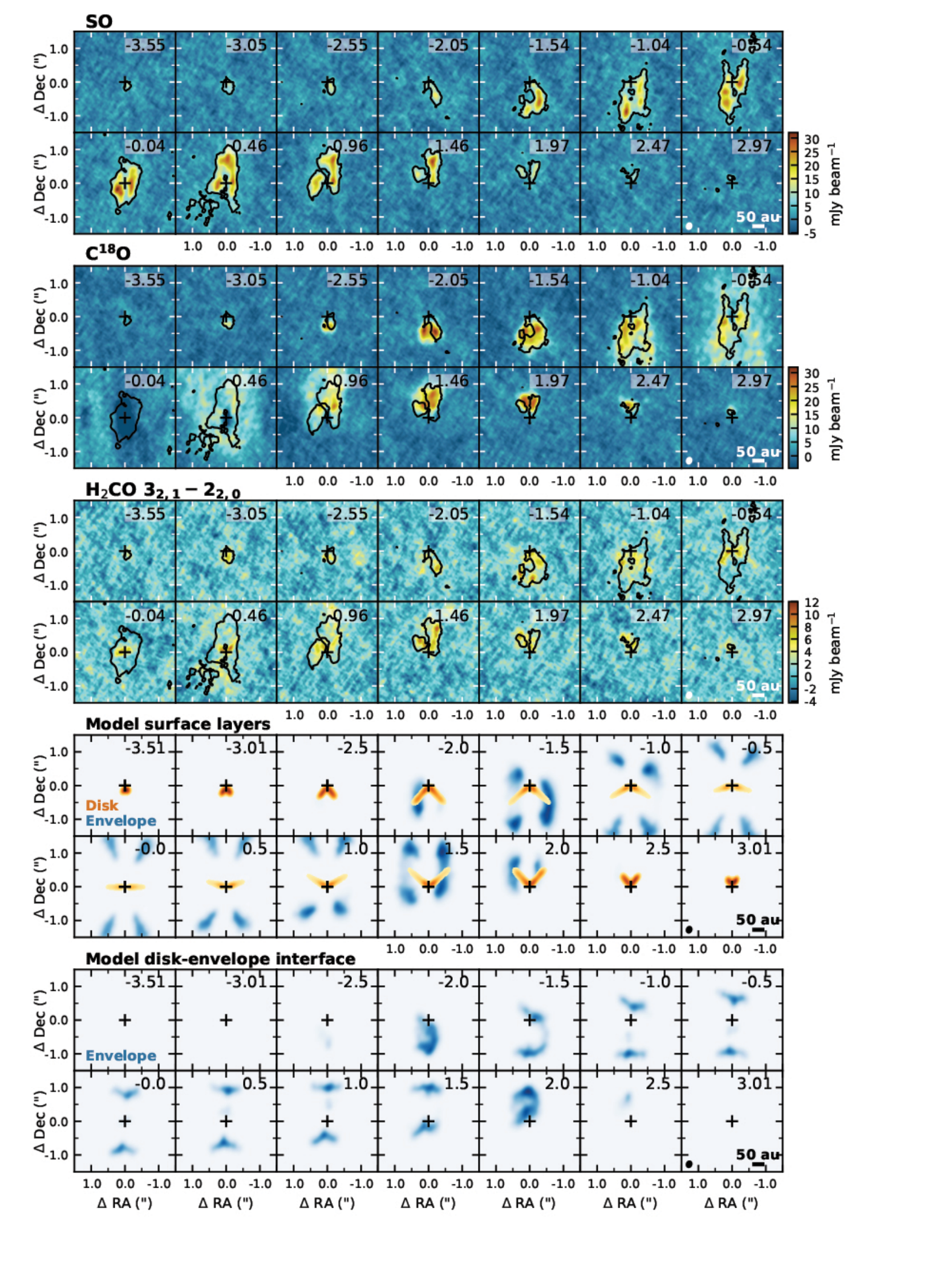}
\caption{Selected velocity channel maps of the 0.167 \kms resolution data cubes for SO (top group of panels), \CeO (second group of panels), and \HtCO $3_{2,1}-2_{2,0}$ (third group of panels). The black contour marks the 3$\sigma$ level of the SO emission in all panels. The fourth and fifth group of panels show velocity channel maps for a model with emission arising in the surface layers of both the disk (orange) and envelope (i.e., along the cavity wall; blue), and for a model with emission arising at the disk--envelope interface, i.e., in the inner envelope (125--150 au), respectively. The black cross marks the source position. The velocity with respect to the systemic velocity is listed in the top right corner of each panel. The beam is shown in the bottom left corner of the right most panels.}
\label{fig:SO_channels}
\end{figure*}

\subsubsection{SO} \label{sec:SO}

Emission from the SO $6_5-5_4$ transition ($E_{\rm{up}}$ = 35 K) is detected over a velocity range of $-$4.05 -- 3.30 \kms (Fig.~\ref{fig:SO_channels_full}), and extends about 1\asec north and south of the source position in the integrated and peak intensity maps (Fig.~\ref{fig:Molecularlines}). At low velocities ($-$0.54 -- 0.46 \kms), weak more extended emission is visible (see Figs.~\ref{fig:12CO_C18O_SO} and~\ref{fig:SO_M8}). These narrow arc-like emission features extend predominantly towards the southwest and are blueshifted, but less extended ($\lesssim$4\asec) X-shaped emission is visible. This large scale emission overlaps with the surface layer of the envelope traced in \CeO. However, the SO emission originates in a more narrow layer in individual velocity channels. Due to the rotating, infalling velocity profile of the envelope, emission from the surface layer has a slightly different spatial location at different velocities, resulting in the double layered feature in the southwest in the peak intensity map (Fig.~\ref{fig:12CO_C18O_SO}, arrow in top right panel). The narrow SO emission feature visible in the southeastern outflow cavity in the peak intensity map (Fig.~\ref{fig:12CO_C18O_SO}, arrow in bottom right panel) is not visible in the \CeO map, but it is formed by small SO emission patches over a velocity range of $-$0.54 -- 0.13 \kms that coincide with the outward moving emission front visible in \thCO and \CeO channels (Figs.~\ref{fig:C18O_channels} and \ref{fig:13CO_channels}). Several more of such SO emission spots coinciding with the outflowing material traced in \thCO and \CeO are visible at higher velocities and larger spatial offsets, especially in the western outflow cavity (Fig.~\ref{fig:SO_M8}). 

The SO emission on 1\asec scales does not resemble the morphology of any of the other lines (Fig.~\ref{fig:Molecularlines}). The north--south emission features east and west of the source position are more parallel to each other than the X-shape seen in \HtCO and do not extend as far east--west as \thCO and \CeO. In addition, the SO emission peaks directly east and west of the source in the integrated intensity map (with the west side being brighter), while \thCO and \CeO clearly peak north and south of source with less emission originating directly to the east and west. From the pv-diagram (Fig.~\ref{fig:pvdiagrams}) it becomes clear that the redshifted emission is brighter than the blueshifted emission. Based on the channel maps (Fig.~\ref{fig:SO_channels}), this is most likely because there is redshifted emission present along the midplane at an offset of $\sim$1\asec, while this is not the case at blueshifted velocities (e.g. 0.46 \kms versus -0.54 \kms). 

The velocity structure as shown in Fig.~\ref{fig:Molecularlines} deviates from pure Keplerian rotation with blueshifted velocities extending north of the source in the west and redshifted velocities extending south in the east. A similar pattern is seen for \thCO, \CeO, \cCtHt, and \HtCO $3_{0,3}-2_{0,2}$ and indicates the presence of envelope emission. The pv-diagrams (Fig.~\ref{fig:pvdiagrams}) show that this envelope component is less extended than for \thCO and \CeO as the SO emission is confined to smaller angular offsets. The SO pv-diagram consists of two components, most clearly distinct at blueshifted velocities, that seem to anti-correlate with the \CeO emission. The first component is a narrow feature at low spatial offsets ($\lesssim$ 0\farcs4) that extends over the entire detected velocity range. A similar, but stronger, feature is also visible for \HtCO that originates predominantly in the disk surface layers, suggesting that this feature traces emission from the disk surface. The second feature is located at lower velocities ($\lesssim$ 2 \kms) and larger spatial offsets and is just offset from the region with bright \CeO emission, suggesting that it originates in the inner envelope or outer disk. 

To look at the spatial origin of the SO emission in more detail we present selected velocity channel maps in Fig.~\ref{fig:SO_channels} (the full velocity range is presented in Fig.~\ref{fig:SO_channels_full}) and overlay the 3$\sigma$ contour of the SO emission on \CeO and \HtCO ($3_{2,1}-2_{2,0}$) channels. At high redshifted velocities (1.13 -- 2.47 \kms), the SO emission displays a V-shape resembling the \HtCO emission and coinciding with the outermost layer of \CeO emission. A similiar pattern is visible at blueshifted velocities ($-$2.38 -- $-$1.88 \kms), although the emission is more asymmetric and dominated by the west side of the disk. This morphology is consistent with the conclusion drawn from the pv-diagrams that part of the SO emission originates in the disk surface layers. 

At lower velocities the emission is predominantly located in two narrow bands extending north--south both east and west of source. Similar features are present in \CeO channels on top of more extended emission. Only at low redshifted velocities (0.29 -- 0.69 \kms) SO emission is visible along the midplane between $\sim$0\farcs5--1\farcs0 north of source, coinciding with the low energy \HtCO ($3_{0,3}-2_{0,2}$) transition. A model with emission originating solely in the surface layers of both the disk and envelope can explain the vertical emission bands at intermediate velocity offsets (e.g., $-$1.54 and 1.46 \kms), but not at velocities close to the systemic velocity (Fig.~\ref{fig:SO_channels}). This could be because there is a disconnect at the disk--envelope interface between the velocity structure in a model with a pure Keplerian disk and CMU envelope. Figure~\ref{fig:SO_channels} also shows results for a model with emission arising solely from the disk--envelope interface. In this scenario, the majority of the emission is expected along the major axis of the system, which does not resemble the observed emission morphology. Overall, the SO emission thus seems to originate in the disk and inner envelope, and in both components the emission seems to originate predominantly in the surface layers.


\section{Analysis and discussion}


\subsection{Continuum sub-structures and inclination} \label{sec:Inclination}

The L1527 disk displays smooth, but asymmetric continuum emission, with the southern side brighter than the northern side along the major axis, and the eastern side brighter than the western side. The image does not show the clumps previously reported in 7 mm VLA images \citep{Nakatani2020}, consistent with more recent higher signal-to-noise VLA images at the same wavelength and spatial resolution \citep{Sheehan2022}. 

The north--south asymmetry was previously observed at 7 mm and 1.3 cm with the VLA, and confirmed through analytic modeling \citep{Sheehan2022}. As discussed by those authors, determining the underlying physical nature of this asymmetry is difficult due to the edge-on nature of the disk, and could depend on where the emission at different wavelengths becomes optically thick. Detailed modeling of multi-wavelength data is therefore required to assess whether there is an enhancement in surface density in the southern part of the disk and if so, whether this is related to for example a vortex, spiral or pressure bump. The east--west asymmetry was not visible in the VLA data, even though they have comparable spatial resolution, potentially because at these longer wavelengths the vertical extent of the disk is smaller and the emission is less optically thick. 

Both asymmetries were recently observed with ALMA in Band 7 (0.87 mm) and Band 3 (3.3 mm), although with the western side brighter in Band 3 \citep{Ohashi2022}. The same Band 3 data was presented by \citet{Nakatani2020}, but they show that the brightness temperature peaks slightly southeast of the source position. We therefore re-imaged the archival Band 3 data with the eDisk data reduction and imaging scripts, which results in an image with the eastern side brighter than the western side (Fig.~\ref{fig:Continuum_3mm}), similar to the Band 6 and 7 images. For a near-edge on disk like L1527, the disk can either be orientated such that eastern side faces us, in which case the emission east of source traces the back side of the disk, that is, the half of the disk furthest away from us along our line of sight, or such that the western side faces us, in which case the emission west of source traces the back or far side of the disk (see e.g., Fig. 6 in \citealt{Oya2015}). An asymmetry along the minor axis of a near-edge on disk can be explained by a vertically extended optically thick dust disk. In this scenario, warmer material is observed toward the back side of the disk because the emission would get optically thick already in the colder outer disk on the near side (e.g., \citealt{Ohashi2022} and Fig. 3 therein, Lin et al. subm., Takakuwa et al. in prep.). The north--south asymmetry appears stronger at 3.3 mm, especially in the east, while the east--west asymmetry is less pronounced at longer wavelengths and only visible in the south. A stronger east--west asymmetry at shorter wavelengths hints that it is indeed due to optically thick emission, as the disk becomes less optically thick at longer wavelengths. However, the resolution of the Band 3 data is slightly lower than the here presented Band 6 data (0\farcs086 $\times$ 0\farcs043 versus 0\farcs056 $\times$ 0\farcs029) and detailed modeling of the continuum emission at multiple wavelengths in the visibility plane is required to confirm the origin of east--west asymmetry. Nonetheless, the images presented here suggest that the continuum presents a coherent picture at different ALMA wavelengths, implying that the eastern side is the back side of the disk under the assumption of optically thick emission.

\begin{figure*}
\centering
\includegraphics[width=\textwidth,trim={0.0cm 17.7cm 0.0cm 0.5cm},clip]{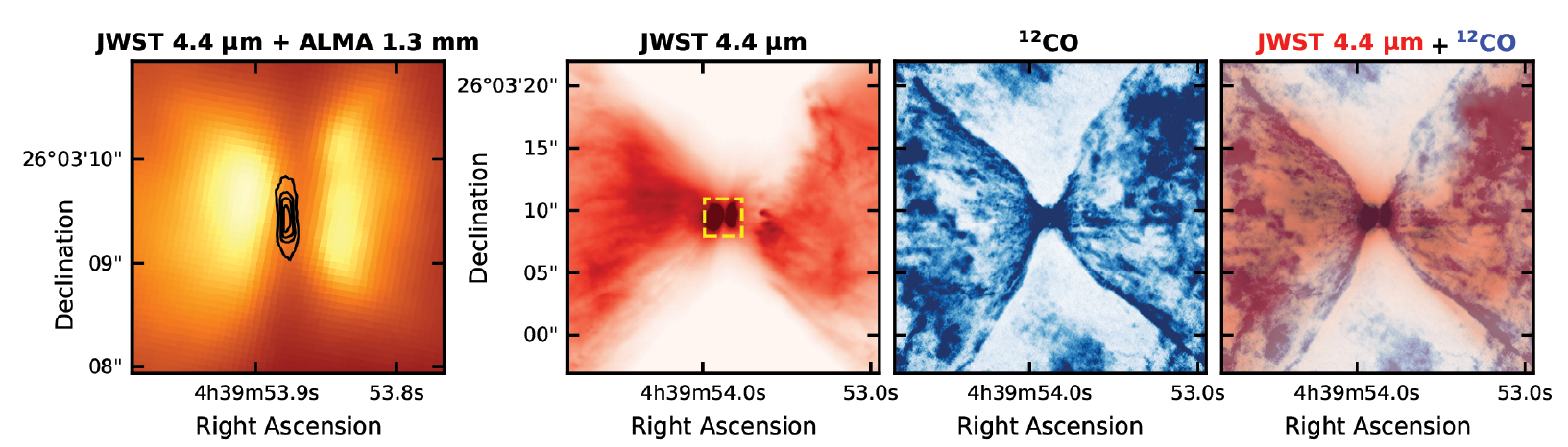}
\caption{Comparison between the ALMA 1.3 mm continuum and \twCO images and the JWST NIRCam 4.4 $\mu$m image. The 4.4 $\mu$m image is shown on different spatial scales in the first and second panel, with the 1.3 mm continuum overlaid in contours in the first panel. The dashed yellow square in the second panel marks the region shown in the first panel. The \twCO peak intensity (moment 8) map is shown in the third panel, and an overlay of the 4.4 $\mu$m image (red) and the \twCO map (blue) is shown in the fourth panel.}
\label{fig:JWST}
\end{figure*}

The brightness asymmetry of the disk previously observed in the Gemini L$^{\prime}$ (3.8 $\mu$m) scattered light image, with the eastern side nearly twice as bright as the western side, also suggests that the eastern side is the side facing us \citep{Tobin2010}. Deconvolution of the Spitzer IRAC image (3.6 $\mu$m) results in the western side being brighter \citep{Velusamy2014}, but the brightness of the cavities in scattered light has been shown to be variable over time \citep{Tobin2008,Cook2019}. The recently released JWST NIRCam image (release id 2022-055, proposal id 2739, PI: K. Pontoppidan\footnote{https://webbtelescope.org/contents/news-releases/2022/news-2022-055}) also suggests that the eastern side is facing us. The JWST 4.4 $\mu$m image (Fig.~\ref{fig:JWST}) is consistent with the Gemini image\footnote{The ALMA continuum seems slightly rotated with respect to the JWST 4.4 $\mu$m image and is not located in the center of the dark lane between the two bright regions east and west of the source position in the 4.4 $\mu$m image. The JWST image aligns with the Spitzer IRAC image presented by \citet{Tobin2008} and there is no rotation evident based on the location of background stars. The east--west misalignment is not due to proper motion as L1527 is moving south, but the displacement is too small to rule out a systematic offset in the NIRCam image using the Spitzer image.} and there is more short wavelength emission (2 $\mu$m) in the eastern cavity, which indicates less extinction and hence that the eastern cavity is the blue-shifted outflow cavity. 

The orientation inferred from the dust continuum and scattered light is at odds with the orientation derived from molecular line emission from the envelope (CS; \citealt{Oya2015}), which suggests that the western side is the back side. In this orientation, the blueshifted envelope emission is stronger in the southwest than in the southeast, while redshifted emission is stronger in the northeast than in the northwest. For a system with the eastern side being the back side, the blueshifted and redshifted envelope components would be strongest in the southeast and northwest, respectively. The velocity patterns of \thCO, \CeO, \cCtHt, \HtCO $3_{0,3}-2_{0,2}$ and SO observed here (Fig.~\ref{fig:Molecularlines}) are consistent with the earlier CS observations and suggest that the western side of the envelope is the far side. 

However, a small inclination toward the west, as suggested by the continuum observations, is consistent with the large scale ($\sim$100\asec) outflow observed in \twCO, which shows blueshifted emission predominantly toward the east and redshifted emission toward the west \citep{Hogerheijde1998}. Consistent with previous observations, on smaller scales as observed here, the \twCO emission displays both blueshifted and redshifted emission on both sides of the source, making it hard to infer the systems inclination. The potential blueshited jet feature in the west (Fig.~\ref{fig:CO_maps}), as well as the blueshifted SiO emission in the west (Fig.~\ref{fig:SiO_12CO}), would suggest an orientation opposite of that based on the large scale outflow. However, in the \twCO velocity channels, weak features are visible at high redshifted velocities in the extension of the potential jet. 

Overall, the continuum emission on small scales ($\lesssim$0\farcs5), the scattered light images, and the large-scale \twCO outflow ($\sim$100\asec) suggest that the eastern side of the system is facing us, while molecular line emission on small scales (SiO jet) and intermediate scales (envelope) are consistent with the opposite orientation of the western side facing us. Differences in orientation based on outflowing material on small and large scales could be due to precession of the outflow, as discussed in more detail by \citet{Oya2015}. The continuum asymmetry is observed at scales of $\sim$0\farcs1, while the orientation derived from the line emission is based on envelope scales. This suggests that there is either a misalignment between the disk and the envelope, or a warp in the inner disk \citep{Cook2019,Sakai2019}. A detailed analysis of the molecular line emission on disk-scales may be able to help constrain the system's orientation. Moreover, an overlay of the \twCO outflow cavity on the JWST NIRCam 4.4 $\mu$m image shows that the asymmetry in the scattered light cavity shapes is not reflected in the \twCO cavity walls, and may thus be the result of shadowing on the northwestern and southeastern cavity wall (Fig.\ref{fig:JWST}). A comprehensive study including the molecular line emission, the multi-wavelength continuum emission, and the scattered light images is therefore required to fully unravel the structure of the system.

\begin{figure*}
\centering
\includegraphics[trim={0.0cm 16cm 0.5cm 0.0cm},clip]{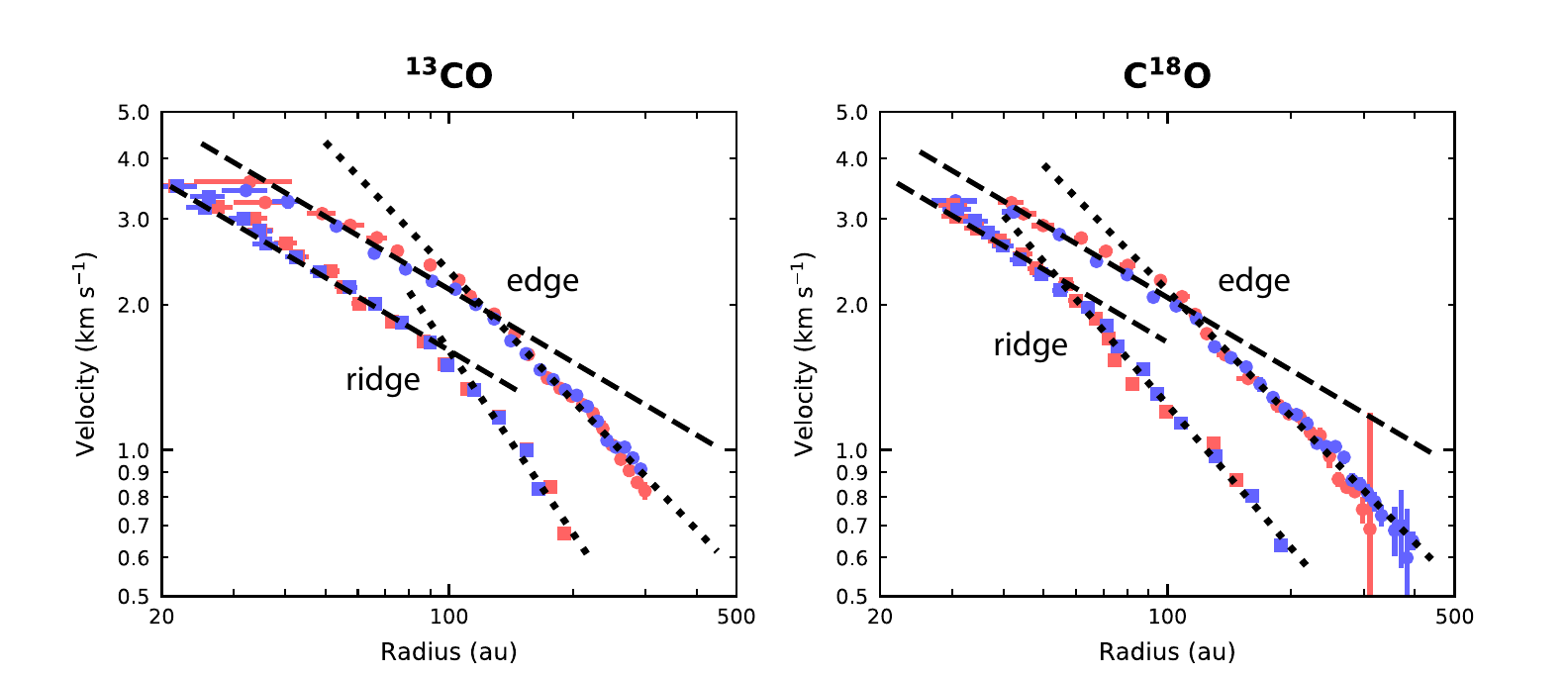}
\caption{Rotation curves derived from the pv-diagrams along the disk major axis (Fig.~\ref{fig:pvdiagrams}) for \thCO (left panel) and \CeO (right panel). Points tracing the outer edge of the pv-diagrams are shown as circles and points tracing the peak of the emission (the ridge) are shown as squares. Blue and red symbols denote blueshifted and redshifted emission, respectively. The dashed lines show power laws with indices, $p_{\mathrm{in}}$, of $-0.5$ (for Keplerian rotation), and the dotted lines show power laws with the best-fit indices, $p_{\mathrm{out}}$, as listed in Table~\ref{tab:pv-analysis}. The radius where the two power laws cross, $r_{\rm{b}}$, is listed in Table~\ref{tab:pv-analysis}. The velocity is with respect to the systemic velocity.}
\label{fig:rotationcurves}
\end{figure*}

\begin{deluxetable*}{@{\extracolsep{4pt}}lcccccccc}
\tablecaption{Results from the pv-diagram analysis. \label{tab:pv-analysis}}
\tablewidth{0pt}
\addtolength{\tabcolsep}{-1pt} 
\tabletypesize{\scriptsize}
\tablehead{
\colhead{} & \colhead{} & \multicolumn{3}{c}{Double power law} & \multicolumn{3}{c}{Inner power law fixed ($p_{\rm{in}}$ = 0.5)} \\
\cline{3-5} \cline{6-8}
\colhead{Method} & \colhead{$v_{\rm{sys}}$ $^a$} & \colhead{$p_{\rm{in}}$ $^b$} & \colhead{$p_{\rm{out}}$ $^c$} & \colhead{$r_{\rm{b}}$ $^d$} & \colhead{$p_{\rm{out}}$ $^c$} & \colhead{$r_{\rm{b}}$ $^d$} & \colhead{$M_\star$ $^e$} \\ [-0.25cm]
\colhead{} & \colhead{(\kms)} & \colhead{} & \colhead{} & \colhead{(au)} & \colhead{} & \colhead{(au)} & \colhead{($M_\sun$)} \\ [-0.45cm]
} 
\startdata 
\thCO edge  & 5.955 $\pm$ 0.005 & 0.43 $\pm$ 0.02 & 0.89 $\pm$ 0.03 & 114 $\pm$ 4 & 0.89 $\pm$ 0.02 & 125 $\pm$ 3 & 0.53 $\pm$ 0.02 \\
\thCO ridge & 6.024 $\pm$ 0.006 & 0.63 $\pm$ 0.01 & 1.57 $\pm$ 0.08 & 131 $\pm$ 3 & 1.25 $\pm$ 0.03 & 100 $\pm$ 2 & 0.29 $\pm$ 0.01 \\
\CeO edge   & 5.958 $\pm$ 0.006 & 0.47 $\pm$ 0.03 & 0.86 $\pm$ 0.04 & 102 $\pm$ 6 & 0.86 $\pm$ 0.02 & 109 $\pm$ 5 & 0.49 $\pm$ 0.03 \\
\CeO ridge  & 5.995 $\pm$ 0.007 & $-$             & $-$             & $-$         & 0.98 $\pm$ 0.02 & 54 $\pm$ 2 & 0.32 $\pm$ 0.02 \\ 
\enddata
\vspace{0.1cm}
\textbf{Notes.} $^a$ Systemic velocity. $^b$ Power-law index of the inner power law. $^c$ Power-law index of the outer power law. $^d$ Transition radius between the inner and out power law. $^e$ Central mass. 
\end{deluxetable*}


\subsection{Dynamical mass and disk radius} \label{sec:DynamicalMass}

The stellar mass and disk radius can be derived by fitting a double power law to the rotation curve \citep[e.g.,][]{Seifried2016,Aso2020,Maret2020}. We use the publicly available Spectral Line Analysis/Modeling (SLAM) code\footnote{\url{https://github.com/jinshisai/SLAM}} to extract the rotation curves from the \thCO and \CeO pv-diagrams \citep{Aso2015, Sai2020} and to do the fitting. The signal-to-noise ratio of the \HtCO and SO emission is not high enough for this analysis and \twCO is dominated by outflow emission. Details of the methods employed by SLAM are described by Ohashi et al. (subm.), but the main steps are as follows. 

First, the (position,velocity) coordinates are determined either for points tracing the outer edge (5$\sigma$ level) of the pv-diagram or for points tracing the peak of the emission (also called the ``ridge'' of the pv-diagram). Next, a double power law is fit to the (position,velocity) coordinates using the Markov Chain Monte Carlo (MCMC) algorithm implemented in the \texttt{emcee} package \citep{Foreman-Mackey2013}. For an infalling rotating envelope conserving angular momentum, the velocity is proportional to $r^{-1}$, while for a Keplerian disk, $v \propto r^{-0.5}$. In the latter case the stellar mass can be derived from $v_{b} = \sqrt{GM_{*} / r_{b}} \sin i$, where $v_b$ is the velocity at the transition radius, $r_b$, between disk and envelope velocity profiles. We first perform the fits with the systemic velocity and inner power-law index as free parameters and confirm that the power-law index for the inner region is indeed $\sim$ $-$0.5. We then fix the systemic velocity to the best-fit value and the inner power-law index to $-$0.5 to derive the dynamical stellar mass. The results are listed in Table~\ref{tab:pv-analysis} and the rotation curves are shown in Fig.~\ref{fig:rotationcurves}.

The dynamical mass derived from the two isotopologues is consistent within 1-2$\sigma$, and is $\sim$0.3 $M_\sun$ using the ridge method and $\sim$0.5 $M_\sun$ using the edge method. This is higher than the 0.2 $M_\sun$ derived by \citet{Tobin2012} from \thCO and  consistent with 0.45 $M_\sun$ derived by \citet{Aso2017} from \CeO, who used the ridge method on lower resolution data. For the edge points, the transition radius found from \thCO and \CeO is also consistent with each other within error bars, while the results from the ridge points are different for the two isotopologues and different from the edge point derived values. The reason is most likely that the ridge is not well defined at this high resolution (see the pv-diagrams in Fig.~\ref{fig:pvdiagrams}). The difference between \thCO and \CeO for the ridge points could also be due to the difference in optical depth between the two isotopologues resulting in both isotopologues tracing a different layer. An average radius of 108 au from the edge points is only slightly larger than the radius derived from the \CeO pv-diagram by \citet{Aso2017} (75 au). \citet{Tobin2012} derived a 125 au disk radius from multi-wavelength continuum modeling, but the 5$\sigma$ contour of the 1.3 mm image extends only out to $\sim$85 au. A detailed analysis of the continuum emission is required to better constrain the dust-disk size, but it seems not very different from the gas-disk size. Overall, the analysis for L1527 seems most robust using the edge method and suggests a stellar mass of $\sim$0.5 $M_\sun$ and a gas-disk radius of $\sim$110 au.


\subsection{Temperature structure} \label{sec:Temperature}

\subsubsection{The disk is too warm for CO freeze out}

On scales of a few arcseconds, both \thCO and \CeO display an X-shaped emission morphology (most clearly visible in the peak intensity maps in Fig.~\ref{fig:CO_maps}), with emission arising from the surface of the envelope along the outflow cavity wall out to larger radii than emission from the midplane. Along the midplane, the emission starts to disappear at offsets of $\sim$2\farcs5 ($\sim$350 au), suggesting that CO starts to freeze out. This is consistent with the analysis of \thCO, \CeO and C$^{17}$O by \citet{vantHoff2018b,vantHoff2020}, which showed that CO was present in the gas phase throughout the entire disk ($\sim$100 au), as well as with the temperature profile derived from modeling the multi-wavelength continuum emission by \citet{Tobin2013} in which the temperature drops below 20 K at a radius of 360 au.  

\begin{figure*}
\centering
\includegraphics[width=\linewidth,trim={0.5cm 19cm 0.5cm 0.4cm},clip]{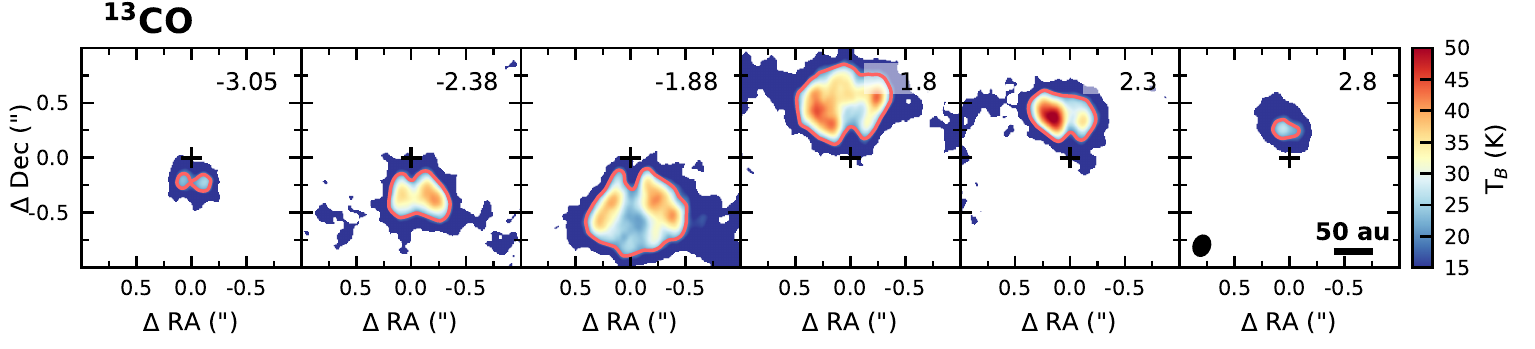}
\caption{Brightness temperature of \thCO in selected 0.167 \kms velocity channels. Only pixels above the 3$\sigma$ level are shown. The red line marks the 20 K contour. The velocity with respect to the systemic velocity is listed in the top right corner of each panel. The beam is shown in the bottom left corner of the right most panel.}
\label{fig:CO_temperature}
\end{figure*}

A more detailed temperature structure can be obtained from the optically thick \thCO brightness temperature as presented in Fig.~\ref{fig:CO_temperature}. For all channels with velocities $\gtrsim |1|$ \kms and with resolved emission, the brightness temperature is higher than 20 K, with temperatures increasing in the surface layers up to $\sim$50 K. The temperature in the radially and vertically most extended regions is $<$20 K, but this is likely due to the emission becoming optically thin (especially in the upper most surface layers) and beam dilution as this region typically fills only half of the beam. Beam dilution is also the reason why temperatures decrease at velocities $\gtrsim |3|$ \kms, where the emission solely originates in the inner half of the disk. Based on the \thCO/\CeO line ratio, the \CeO emission is only optically thick in the midplane at angular offsets $\lesssim$ 0\farcs5. Consequently, the midplane brightness temperatures are consistent with those for \thCO, while the brightness temperature in the surface layers is lower ($\sim$30--35 K). These results are also in agreement with the disk midplane temperature profile derived by \citet{vantHoff2018b} from  observations with lower resolution and sensitivity.

\begin{figure*}
\centering
\includegraphics[width=\linewidth,trim={0.5cm 16.2cm 0.5cm 0.4cm},clip]{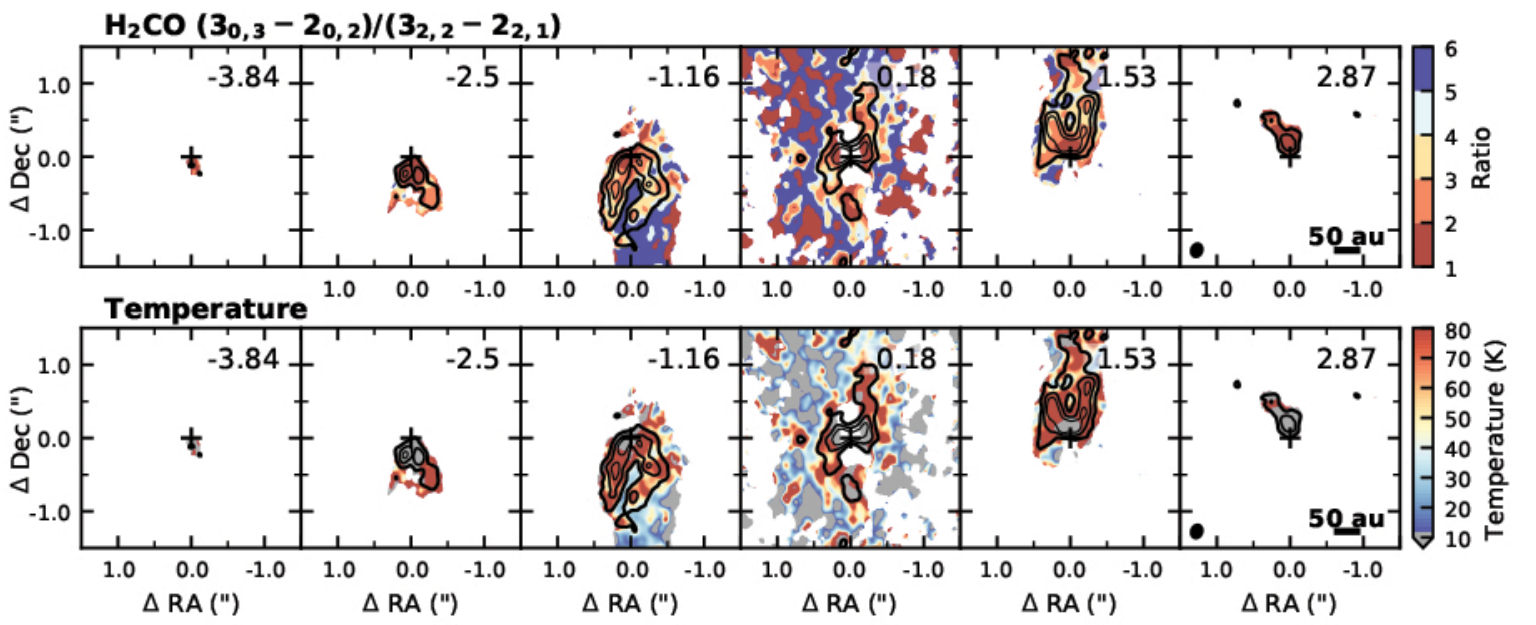}
\caption{Line ratio of the \HtCO $3_{0,3}-2_{0,2}$ and $3_{2,2}-2_{2,1}$ transitions in the observed velocity range (top panels) and the corresponding temperature for optically thin emission (bottom panels). Ratios lower than 2 indicate optically thick emission and those regions are shown in grey in the temperature maps (bottom panels). Only pixels with a $> 3\sigma$ detection of the $3_{0,3}-2_{0,2}$ transition are shown, and the contours depict the 3, 6 and 9$\sigma$ levels of the $3_{2,2}-2_{2,1}$ transition. The color scale is such that blue indicates low temperatures (corresponding to high ratios) and red indicates high temperatures (corresponding to low ratios). The black cross marks the source position. The velocity with respect to the systemic velocity is listed in the top right corner of each panel. The beam is shown in the bottom left corner of the right most panels.}
\label{fig:H2CO_temperature}
\end{figure*}

The temperature structure can also be assessed using \HtCO. The resolution is high enough to spatially resolve a decrease in \HtCO toward the midplane, as observed before for the edge-on young disk IRAS 04302 \citep{Podio2020,vantHoff2020}. Assuming this is due to freeze out, the base of the V-shape provides an estimate of the \HtCO snowline at $\sim$70 K \citep{Noble2012,Fedoseev2015}. The tip of the V is unresolved, suggesting that the snowline is at a radius less than 24 au. In addition, the ratios of \HtCO lines are good probes of temperature \citep[e.g.,][]{Mangum1993}. Fig.~\ref{fig:H2CO_temperature} (top row) presents the $3_{0,3}-2_{0,2}$/$3_{2,2}-2_{2,1}$ line ratio per channel, and shows that the ratio is $\lesssim$3 in all pixels with a $> 3\sigma$ detection of the weaker $3_{2,2}-2_{2,1}$ transition.

The $3_{0,3}-2_{0,2}$/$3_{2,2}-2_{2,1}$ ratio as observed here is particularly sensitive to temperatures $\lesssim$ 90 K (see Fig.~\ref{fig:H2CO_ratio}). For optically thin emission in Local Thermodynamic Equilibrium (LTE), the ratio drops from $\sim$20 to $\sim$5 for temperatures between 20 and 40 K. Around 90 K the ratio has dropped to $\sim$3, and the ratio remains higher than 2 for temperatures up to 300 K. If the emission is optically thick, the ratio drops below 2. The observed line ratio is $>2$, except in the inner $\sim$0\farcs3, where the emission thus becomes optically thick. A conservative estimate of the uncertainty on the line ratio in the optically thin regions (ratio $>$ 2) with a $> 6\sigma$ detection of the $3_{2,2}-2_{2,1}$ transition is then 18\% when propagating the individual rms noise levels of the observations. This means that for line ratios of $\sim$3, the temperature is at least 60 K, but the upper limit is not well constrained. As can be seen from Fig.~\ref{fig:H2CO_temperature}, the temperature in the regions where both \HtCO lines are detected is then at least 60 K. At velocity offsets of $-$1.16 and 1.53 \kms, no $3_{2,2}-2_{2,1}$ emission is detected from the midplane and the ratio becomes $\gtrsim$4, suggesting temperatures below $\sim$40 K. In the inner envelope midplane where both transitions are detected (angular offsets of $\sim$0\farcs75-1\farcs0) temperatures of at least 60 K are found. A similar temperature was derived in this region from a non-LTE large-velocity-gradient (LVG) analysis of SO by \citet{Ohashi2014,Sakai2014a}.

The temperatures derived from the \HtCO line ratio are $\sim$10--20 K higher than the \thCO brightness temperature. This difference is likely caused by two dominant effects. First, the brightness temperature from \thCO traces the radius where the emission gets optically thick. This is not necessarily the same as the observed angular offset, which means that colder material further out may be projected closer to the star. Second, the brightness temperature of the \thCO emission gets smeared out over the size of the beam, while the effects of beam dilution cancel out for the \HtCO line ratio. Since both effects work to lower the observed brightness temperature of \thCO, the physical temperature is likely closer to that measured with \HtCO. In addition, while the brightness temperature of the continuum drops steeply at radii $\gtrsim$0\farcs3, continuum subtraction may have lowered the \thCO brightness temperature a little. A higher midplane temperature than derived from the \thCO brightness temperature would be consistent with the analysis by \citet{vantHoff2018b} that showed that the temperature profile as derived by \citet{Tobin2013} from the multi-wavelength continuum emission needed to be increased by at least 30\% to reproduce the observations. Such a warm temperature profile was also adopted by \citet{Aso2017} for analysis of \CeO observations of L1527.

Finally, if the dust is optically thick, as suggested by the brightness asymmetry along the minor axis, the continuum brightness temperature provides a third temperature probe. The continuum brightness temperature is $\sim$40 K out to $\sim$25 au, after which it steeply drops to temperatures below $\sim$20 K (Fig.~\ref{fig:Continuum}), probably because the emission becomes optically thin. A midplane temperature of $\sim$40 K at 25 au is consistent with the \thCO brightness temperatures at larger radii and the temperature profile derived by \citet{vantHoff2018b}. However, the continuum brightness temperature suffers from the same effects as the \thCO brightness temperature, so it may also underestimate the real temperature. Overall, all results point to the L1527 disk being warm, with midplane temperatures too high for CO freeze out throughout the disk (20--40 K), but low enough for H$_2$CO freeze out outside $\sim$24 au.  

Moreover, the similarity between the temperature derived from the continuum and the molecular lines suggests that the grains do not scatter efficiently, as this would decrease the brightness temperature \citep{Birnstiel2018}. Since grains only scatter appreciably when the size of the grain is comparable to the observing wavelength, this suggests that L1527 mainly contains grains much smaller than $\sim$1 mm. The lack of large grains is consistent with the vertical extent of the dust disk which indicates that the grains have not yet settled.


\subsubsection{A potential temperature enhancement around the disk--envelope interface}\label{sec:Tincrease}

At intermediate redshifted velocities (e.g., the 1.8 \kms channel shown in Fig.~\ref{fig:CO_temperature}), the \thCO brightness temperature increases from $\sim$25 K to $\sim$35 K at angular offsets $\gtrsim$ 0\farcs5 (1$\sigma$ = 2 K). A similar effect is derived from the \HtCO line ratio, but at both redshifted and blueshifted velocities. This suggests that there may be an increase in temperature around the disk--envelope interface. Such a rise was invoked by \citet{Sakai2014a} based on SO observations. Temperatures of $\sim$60 K and $\sim$200 K were derived from SO emission at angular offsets between $\sim$0\farcs6--1\farcs0 (\citealt{Ohashi2014,Sakai2014a,Sakai2017}, resp.), but these studies did not have high enough spatial resolution and signal-to-noise ratio to derive a temperature at smaller offsets. However, there is no increase in \thCO brightness temperature at blueshifted velocities, so several factors have to be taken into account before a temperature enhancement can be concluded from the observations presented here.

Relating the brightness temperature observed at a certain position in a certain velocity channel to a physical location is non-trivial, because it depends on where in the system the emission becomes optically thick (see e.g., Fig.~6 in \citealt{vantHoff2018b}). For example, if the emission along a line of sight close to the protostar becomes optically thick in the outer envelope, the brightness temperature at a small angular offset will reflect the temperature in the cold outer envelope instead of in the warm disk. Since redshifted emission originates in the rotating-infalling envelope in front of the disk, and blueshifted emission in the envelope behind the disk, a comparison between redshifted and blueshifted velocity channels can help constrain the origin of the emission. If the emission becomes optically thick in the disk, the redshifted and blueshifted channels should be similar. However, this is inconsistent with the observations. The two simplest possible scenarios are then that either the \thCO emission is just optically thick ($\tau \gtrsim$ 1), or that the \thCO is very optically thick ($\tau >>$1). 

The first scenario ($\tau \gtrsim$ 1) could explain the observations if the \thCO emission becomes optically thick in the envelope behind the disk at blueshifted velocities, and in the disk at redshifted emission. In other words, if the disk itself does not contain enough CO to make the emission optically thick. The observations would then suggest that the envelope midplane (seen at blueshifted velocities) is colder than the outer disk midplane (seen at redshifted velocities). A potential issue with this scenario is that it may not be able to reproduce the (marginally) optically thick \CeO emission.  

The second scenario ($\tau >>$1) may explain the observations if the \thCO emission already becomes optically thick in the envelope at redshifted velocities and in the disk at blueshifted velocities. In this case there is enough material in both the envelope and the disk to make the emission optically thick. This would then suggest that the disk midplane (seen at blueshifted velocities) is colder than the inner envelope (seen at redshifted velocities). If the \CeO emission does not become optically thick yet in the envelope (at redshifted velocities), this temperature increase would not be visible in \CeO, consistent with the observations. A potential inconsistency is that you would expect to observe warm emission from the inner envelope at angular offsets larger than the extent of the disk at both blueshifted and redshifted velocities. The fact that this is not observed, suggests then that either the temperature or the optical depth drops quickly in the inner envelope. However, a rapid decrease in temperature is inconsistent with the temperature derived from \HtCO, which shows $\sim$60 K out to $\sim$1\farcs1--1\farcs4 ($\sim$150--200 au projected from the source), and a rapid decrease in optical depth is inconsistent with the observed \thCO/\CeO line ratio, which suggests \thCO is optically thick out to $\sim$1\farcs1--1\farcs4 as well.  

An alternative explanation for the \thCO observations is that there is a north--south asymmetry in the temperature, but this would be opposite to the north--south asymmetry observed for the continuum emission and not clearly visible in \HtCO. In conclusion, both the \thCO and \HtCO show potential evidence of an increase in the temperature in the outer disk and/or inner envelope. However, observations of \HtCO at higher spectral resolution and for higher energy transitions are required to better constrain the temperature. Moreover, detailed radiative transfer modeling is required to derive a temperature structure for the disk and envelope from the observed emission and to confirm whether an increase in temperature occurs at the disk--envelope interface.

\begin{figure}
\centering
\includegraphics[width=\linewidth,trim={0cm 0cm 0cm 0cm},clip]{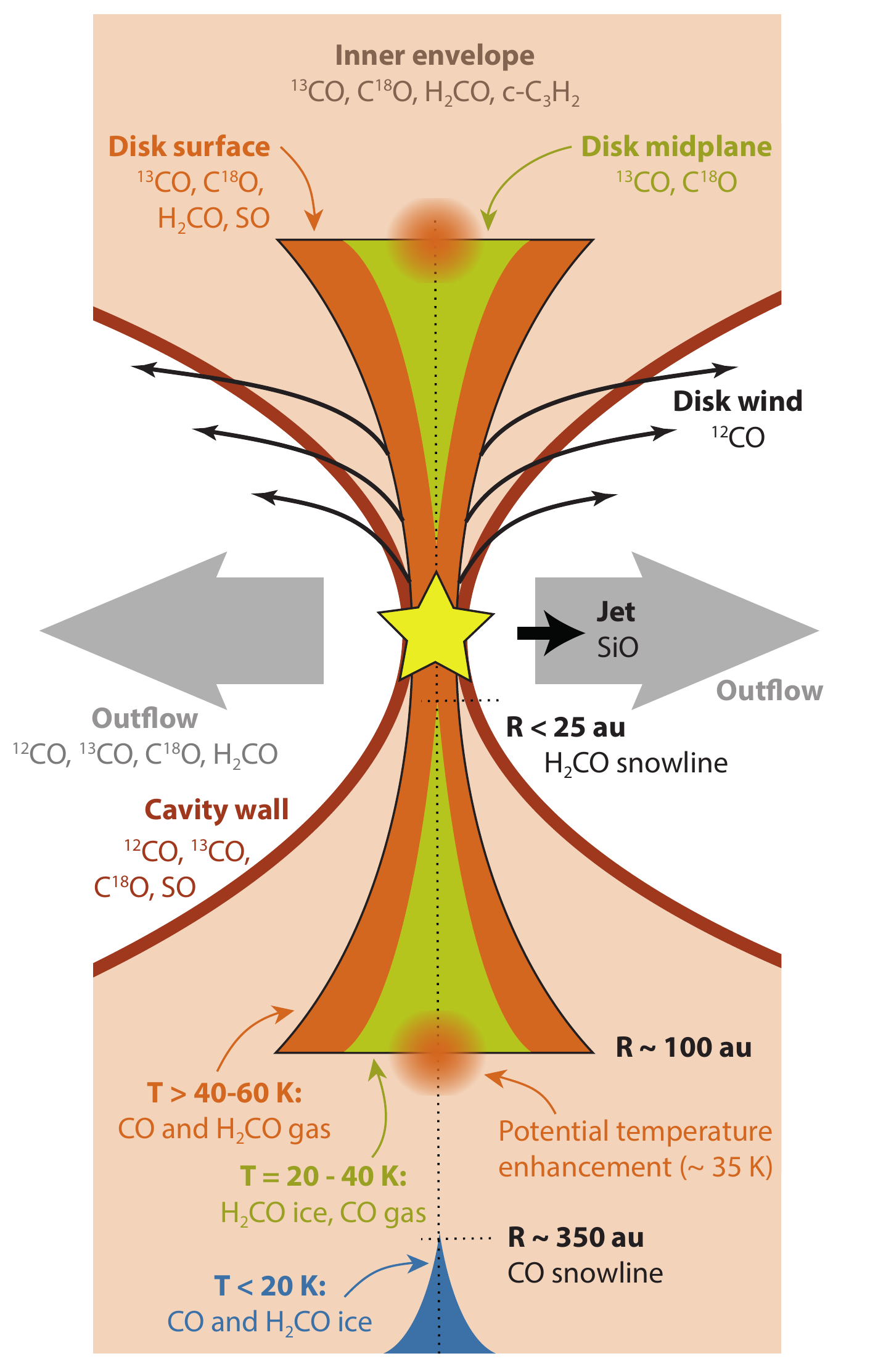}
\caption{Schematic overview of the chemical and temperature structure of the inner few hundred au of the L1527 protostellar system based on the observations presented here. The potential temperature enhancement is seen in both the north and south in \HtCO, but only in the north in \thCO.}
\label{fig:Cartoon}
\end{figure}


\subsection{Molecular structure} \label{sec:MolecularStructure}

A cartoon illustrating which components of the L1527 protostellar system are traced by which molecules is presented in Fig.~\ref{fig:Cartoon}. Outflowing material is visible in \twCO, \thCO, \CeO, and \HtCO $3_{0,3}-2_{0,2}$ with \twCO tracing the highest velocity gas. A blueshifted jet is visible in the western outflow cavity in SiO and potentially \twCO. The SiO jet is compact and located close to the protostar (peak position is 0\farcs08 offset from source), while the \twCO jet is further away from the source ($\sim$3--6\asec) and extended in the east--west direction. The outflow cavity wall is clearly outlined in \twCO, but also faintly in \thCO (at small scales), \CeO and SO. The large scale envelope is visible in \cCtHt and weakly in DCN. Emission from the inner envelope is most strongly visible in \thCO. In addition, while weak envelope emission is seen in \HtCO $3_{0,3}-2_{0,2}$, all \HtCO transitions display bright emission in the inner envelope midplane. The disk is traced by \thCO, \CeO and \HtCO, where \HtCO only originates in the disk surface layers. SO also traces surface layers in the disk and inner envelope, that is, in layers along the outflow cavity wall. The physical and/or chemical reason behind the different emission morphologies is further discussed in the following sections. 

\subsubsection{Outflowing material}

\begin{figure}
\centering
\includegraphics[width=\linewidth,trim={0cm 6cm 7.5cm 0.8cm},clip]{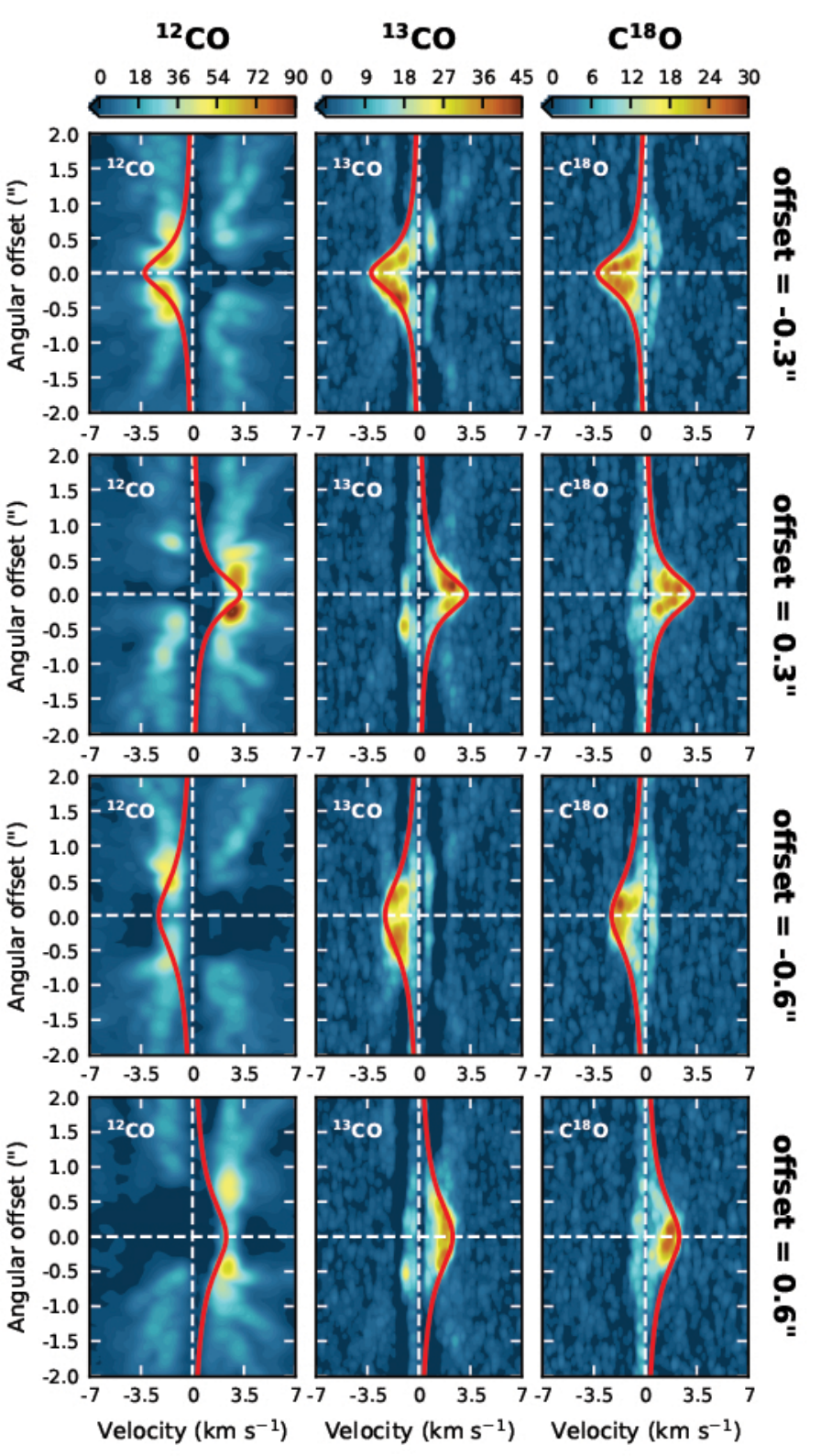}
\caption{Position-velocity diagrams of the CO isotopologues (different columns) extracted parallel to the disk minor axis (averaged over the size of the beam) at offsets of $\pm$0\farcs3 (first and second row) and $\pm$0\farcs6 (third and fourth row) along the major axis (i.e., north and south of the source position). The horizontal dotted line marks the disk midplane, with larger angular offsets corresponding to the disk surface layers. The vertical dotted line marks the systemic velocity, which is shifted to 0 \kms. The red curve shows the expected Keplerian velocity for a 0.5 $M_{\sun}$ star (based on the pv-diagram analysis described in Sect.~\ref{sec:DynamicalMass}) at each radial offset, $r$, as function of disk height, $z$: $V_{\rm{Keplerian}}=\sqrt{\frac{GMr^2}{(r^2+z^2)^{3/2}}}$ \citep[e.g.][]{Rosenfeld2013}.}
\label{fig:pvCO}
\end{figure}

All three CO isotopologues display a moving front of emission in the outflow cavity with material at higher velocities located further away from the star (marked with dotted lines in the channel maps in Figs.~\ref{fig:12CO_channels}-\ref{fig:C18O_channels}). This is in agreement with both a jet/bow shock driven outflow as well as a wind-driven outflow \citep[e.g.,][and references therein]{Lee2000,Arce2007}. The velocity maps in Fig.~\ref{fig:Molecularlines} show that the \twCO emission has higher velocities along the outflow cavity walls close to source ($\lesssim$1\asec) compared to the disk emission traced by \thCO and \CeO, suggesting that there may be a disk wind in this system. The velocity structure can be shown more clearly in pv-diagrams obtained parallel to the disk minor axis at different offsets along the major axis, that is, north and south of the source position (Fig.~\ref{fig:pvCO}). \thCO and \CeO show Keplerian emission on one side of the diagram and a low-velocity contribution from the envelope in the other half. In contrast, \twCO shows extended emission in all four quadrants but peaks near the \thCO and \CeO disk emission. However, the bright \twCO emission has velocities higher than the expected Keplerian velocity. The fainter extended \twCO emission may be explained by material moving away in a conical or parabolic shape. Because the outflow is in the plane of the sky, the near side of this shell would move toward us and the far side would move away from us, resulting in both redshifted and blueshifted emission. This poloidal expansion coupled with the super-Keplerian rotation suggests that \twCO may trace a rotating, expanding disk wind. A more detailed analysis is left for future work. 

The outward moving front is present in both outflow cavities for \thCO and \CeO, while \twCO only shows this in the eastern cavity (Figs.~\ref{fig:12CO_channels}-\ref{fig:C18O_channels}, but outside of the angular range shown in the pv-diagrams in Fig.~\ref{fig:pvCO}). Instead, there is an indication of a blueshifted \twCO jet $\sim$3--6\asec ($\sim$ 400--800 au) off source (Figs.~\ref{fig:CO_maps} and \ref{fig:12CO_channels}). However, the velocity is lower than the velocity of the SiO jet closer to source ($\sim$ 11 au). \twCO emission has been observed before with CARMA ($\sim$3\asec resolution) and ALMA (0\farcs8 resolution) in the $J=1-0$ transition \citep{Flores-Rivera2021}, which displays a similar outflow morphology as the $J=2-1$ transition presented here. \citet{Flores-Rivera2021} referred to the narrow `neck' between the two outflow cavities (see Figs.~\ref{fig:Molecularlines}, \ref{fig:JWST}, \ref{fig:12CO_channels}, fourth row) as a jet extending out to $\sim$75 au. However, the channel maps in Fig.~\ref{fig:12CO_channels} (fourth row) suggest a widening of the outflow opening angle with the north-south extension (making up the `neck') originating from emission cospatial with the disk and/or inner envelope (potentially tracing a wind) with redshifted emission north of the source and blueshifted emission south of the source. At larger scales (Fig.~\ref{fig:12CO_channels}), there is a hint of a redshifted jet-like emission feature along the line of the blueshifted jet at larger offsets ($\sim$8--12\asec). This may signal that the jet has precessed. Precession may also explain the kinks seen in the outflow cavity walls (marked by white arrows in Fig.~\ref{fig:CO_maps}). Alternatively, this could be due to widening of the outflow opening angle over time, or due to inhomogeneity of the surrounding medium.

\subsubsection{Disk versus envelope}

Studying whether changes occur in the composition of the volatile material as it transitions from the envelope to the disk is crucial for understanding the origin of chemical complexity in planet-forming material. Moreover, identifying molecular species or transitions that exclusively trace the disk environment would aid in identifying and studying the conditions in young disks. In recent years, chemical changes across the disk--envelope interface have been observed for multiple sources (e.g., L1527, L1489, IRAS 16293 A, IRAS 16293 B, IRAS 15398; \citealt{Sakai2014a,Yen2014,Oya2016,Oya2018,Okoda2018}, resp.), but so far no molecule or transition has been uniformly identified that reliably traces the disk or the disk--envelope interface. 

In the observations presented here, \cCtHt emission originates solely in the envelope. The \cCtHt observations are consistent with previous ALMA observations of different transitions \citep{Sakai2014a} and the distribution of C$_2$H \citep{Sakai2014b}. These hydrocarbons are often associated with UV irradiation and therefore typically observed in photo-dissociation regions (PDRs; e.g., \citealt{vanderWiel2009,Guzman2015}) and along outflow cavity walls, especially in Class 0 sources \citep[e.g.,][]{Murillo2018,Tychoniec2021}. At the spectral resolution of 1.34 \kms, strong \cCtHt emission is only detected in three 1.34 \kms velocity channels ($-$1.16 - 1.53 \kms), but weak (3$\sigma$) emission is detected at $-$2.5 \kms for the blended $6_{0,6}-5_{1,5}$ and $6_{1,6}-5_{0,5}$ transitions, which is at a higher velocity offset than detected before ($-$2.2 \kms; \citealt{Sakai2014a}). Higher spectral resolution observations are required to constrain the distribution in the envelope, and higher sensitivity observations are needed to establish whether \cCtHt is really absent in the disk. The presence or absence of \cCtHt in the disk is interesting as bright hydrocarbon emission in Class II protoplanetary disks is associated with high C/O ratios ($>$ 1; \citealt{Bergin2016,Miotello2019}). A comparison between embedded and mature disks could thus provide information about the chemical evolution during the disk stage.

The \HtCO transitions observed here predominantly originate in the disk surface layers, where the temperature as determined from the line ratio is $\gtrsim$60 K. The higher energy transitions ($3_{2,1}-2_{2,0}$ and $3_{2,2}-2_{2,1}$) are not detected in the midplane, while weak emission from the lower energy transition ($3_{0,3}-2_{0,2}$) is present. The $3_{0,3}-2_{0,2}$ flux is expected to increase at lower temperatures (for a given column density), so this is not an excitation effect. The observed distribution could instead be due to freeze out of \HtCO in the midplane as the freeze-out temperature is $\sim$70 K \citep{Noble2012,Fedoseev2015}. A low residual midplane abundance could be due to gas-phase formation of \HtCO or nonthermal desorption \citep[e.g.,][]{Aikawa2002,Loomis2015,Oberg2017,TerwisschavanScheltinga2021}. Alternatively, the $3_{0,3}-2_{0,2}$ emission along the midplane originates in the envelope that is visible between the disk surface layers. These results substantiate the analysis of the temperature profile of the young disk IRAS 04302 based on a similar V-shaped emission pattern of one \HtCO transition toward that source \citep{vantHoff2020}. The increase in \HtCO intensity in the inner envelope (or outer disk, as detailed modeling is required to determine the exact location) could then be due to a lower freeze-out temperature at lower densities or an increase in temperature at the disk--envelope interface (as discussed in Sect.~\ref{sec:Tincrease}). 

SO has been suggested to be enhanced at the centrifugal barrier in L1527, due to elevated temperatures in this region caused by an accretion shock \citep[e.g.,][]{Sakai2014a,Sakai2017}. While the concept of a centrifugal barrier is not supported by hydrodynamic simulations, the physics and dynamics at the disk--envelope interface are complex \citep{Jones2022,Shariff2022}, and models show that SO can be enhanced if a shock occurs \citep{Aota2015,Miura2017,vanGelder2021}. Consistent with previous observations, SO emission is strong at scales $\lesssim$1\asec, and in the region of the pv-diagram consistent with the outer disk and/or inner envelope. However, a careful examination of the individual velocity channels suggests that the emission is coming from surface layers of the envelope (i.e., along the cavity wall) rather than the midplane. This is, for example, clearly visible at $-$1.54 \kms, where \HtCO $3_{2,1}-2_{2,0}$ displays emission from the disk surface layers and the envelope midplane, while the SO emission surrounds the \HtCO envelope midplane emission (Fig.~\ref{fig:SO_channels}, but see also Fig.~\ref{fig:H2CO_channels}). The SO emission pattern is also inconsistent with a simple model with emission from the inner envelope (Fig.~\ref{fig:SO_channels}, fifth set of panels), and shows better agreement with a model where the emission solely originates in the surface layers of the disk and envelope (Fig.~\ref{fig:SO_channels}, fourth set of panels). 

Based on the \HtCO line ratio, the temperature in the disk surface layers is $\gtrsim$60 K, so the origin of the SO emission could be thermal sublimation of SO ice as the freeze out temperature of SO is $\sim$40--60 K \citep{Hasegawa1993,Garrod2006}. Thermal sublimation may also explain why there is some SO emission present along the midplane at $\sim$0\farcs7 at low redshifted velocities (0.46-0.96 \kms), as the \thCO brightness temperature is increased in this region. However, the increase in \thCO brightness temperature is visible at higher velocities ($\sim$1.8 \kms; Fig.~\ref{fig:CO_temperature}), while there is no SO emission along the midplane at those velocities. Alternatively, SO may trace UV irradiated environments, either because of UV heating of the gas causing SO ice to desorb, or because its formation becomes possible in these regions through photodissociation of H$_2$O. It is possible to form SO in shocks, which may occur at the disk--envelope interface and/or along the outflow cavity wall, but this still requires the presence of a UV field \citep{vanGelder2021}. If the distribution of SO is set by the UV field, this may naturally lead to an enhancement in the inner envelope as UV radiation may penetrate deeper into the envelope than into the disk. However, a detailed study is required as the disk could shadow parts of the inner envelope. An SO distribution along the outflow cavity wall was also inferred for the embedded disk TMC1A \citep{Harsono2021}. Observations of multiple SO transitions are needed to derive the temperature of the emitting gas and detailed modeling will be required to fully constrain the spatial and physical/chemical origin of the SO emission.


\section{Conclusions} \label{sec:Conclusions}

We have presented high resolution (0\farcs06--0\farcs17 or 8--24 au) ALMA observations (taken as part of the Large Program eDisk) of the 1.3 mm continuum and molecular line emission toward the Class 0 protostar L1527, and provided a qualitative description of the different emission morphologies and their potential underlying physical and/or chemical conditions. The main conclusions are summarized below: 

\begin{itemize}

\item The continuum emission is smooth, but asymmetric along both the major (north--south) and minor axis (east--west), with emission being brighter in the south and east. The flaring nature of the disk and the comparable brightness temperature of the dust and \thCO emission suggest that the grains have not yet grown beyond $\sim$1 mm and settled to the midplane.    

\item Although the disk is viewed nearly edge-on, there is evidence of misalignment between different components as the continuum asymmetry along the minor axis (on scales of $\lesssim$0\farcs05) and the large-scale outflow ($\sim$100\asec) suggest that the east side of the system is the far side, while the SiO jet ($\lesssim$0\farcs1) and envelope emission (a few arcsec) suggest that the west side is the far side. 

\item Different molecules trace different components of the protostellar system. Outflowing material is most clearly visible in \twCO, but also in \thCO, \CeO and \HtCO, while SiO traces a compact jet in the western outflow cavity. The outflow cavity wall is also visible in SO. Super-Keplerian \twCO emission in the inner $\sim$1\asec may trace a disk wind. \cCtHt and DCN only show emission from the envelope, although higher sensitivity is required to rule out their presence in the disk. \thCO, \CeO, \HtCO and SO trace the disk and inner envelope, with \HtCO emission arising predominantly in the disk surface layers. SO emission is dominated by the envelope surface along the outflow cavity wall and the disk component originates in the surface layers. This suggests that SO may be tracing UV irradiated regions. 

\item Analysis of the pv-diagrams of \thCO and \CeO results in a $\sim$100 au Keplerian rotating disk around a $\sim$0.5 $M_{\sun}$ star. 

\item The disk is warm, with temperatures of 20--40 K throughout the midplane and $\gtrsim$50--60 K in the surface layers. The disk is therefore too warm for CO freeze out, which occurs at a midplane radius of $\sim$350 au in the envelope, while \HtCO is frozen out in the midplane down to $\lesssim$25 au. 

\item The \thCO brightness temperature and \HtCO line ratio suggest a potential temperature increase around the disk--envelope interface. However, this increase is only seen at redshifted velocities for \thCO and more detailed modeling is required to determine whether the higher temperature occurs in the outer disk or inner envelope, and whether it is associated with an accretion shock. 

\end{itemize}

High resolution observations of multi-wavelength continuum and multiple molecular species are required to study young disks. Given the complex nature of molecular line emission from a disk--envelope system, these observations need to be combined with source-specific radiative transfer modeling to provide a detailed picture of the physical and chemical structure. The eDisk observations and the work presented here highlight the potential of such studies, and a comparison between the ALMA and JWST NIRCam images already shows the synergy between the two observatories.



\acknowledgments 

We thank the referee for carefully reading the manuscript. This paper makes use of the following ALMA data: ADS/JAO.ALMA\#2019.1.00261.L, ADS/JAO.ALMA\#2019.A.00034.S and 

ADS/JAO.ALMA\#2017.1.00509.S. ALMA is a partnership of ESO (representing its member states), NSF (USA) and NINS (Japan), together with NRC (Canada), MOST and ASIAA (Taiwan), and KASI (Republic of Korea), in cooperation with the Republic of Chile. The Joint ALMA Observatory is operated by ESO, AUI/NRAO and NAOJ. The National Radio Astronomy Observatory is a facility of the National Science Foundation operated under cooperative agreement by Associated Universities, Inc. M.L.R.H acknowledges support from the Michigan Society of Fellows. J.J.T. acknowledges support from NASA XRP 80NSSC22K1159. Z.-Y.L. is supported in part by NASA NSSC20K0533 and NSF AST-1910106. N.O. acknowledges support from National Science and Technology Council (NSTC) in Taiwan through the grants NSTC 109-2112-M-001-051 and 110-2112-M-001-031. J.K.J., S.G., and R.S. acknowledge support from the Independent Research Fund Denmark (grant No. 0135-00123B). Z.-Y.D.L. acknowledges support from NASA 80NSSC18K1095, the Jefferson Scholars Foundation, the NRAO ALMA Student Observing Support (SOS) SOSPA8-003, the Achievements Rewards for College Scientists (ARCS) Foundation Washington Chapter, the Virginia Space Grant Consortium (VSGC), and UVA research computing (RIVANNA). Y.A. acknowledges support by NAOJ ALMA Scientific Research Grant code 2019-13B, Grant-in-Aid for Scientific Research (S) 18H05222, and Grant-in-Aid for Transformative Research Areas (A) 20H05844 and 20H05847. I.d.G. acknowledges support from grant PID2020-114461GB-I00, funded by MCIN/AEI/10.13039/501100011033. P.M.K. acknowledges support from NSTC 108-2112- M-001-012, NSTC 109-2112-M-001-022 and NSTC 110-2112-M-001-057. W.K. was supported by the National Research Foundation of Korea (NRF) grant funded by the Korea government (MSIT) (NRF-2021R1F1A1061794). C.W.L. is supported by the Basic Science Research Program through the National Research Foundation of Korea (NRF) funded by the Ministry of Education, Science and Technology (NRF- 2019R1A2C1010851), and by the Korea Astronomy and Space Science Institute grant funded by the Korea government (MSIT; Project No. 2022-1-840-05). J.E.L. was supported by the National Research Foundation of Korea (NRF) grant funded by the Korean government (MSIT) (grant number 2021R1A2C1011718). L.W.L. acknowledges support from NSF AST-2108794. S.N. acknowledges support from the National Science Foundation through the Graduate Research Fellowship Program under Grant No. 2236415. Any opinions, findings, and conclusions or recommendations expressed in this material are those of the authors and do not necessarily reflect the views of the National Science Foundation. S.T. is supported by JSPS KAKENHI Grant Numbers 21H00048 and 21H04495. This work was supported by NAOJ ALMA Scientific Research Grant Code 2022-20A. S.P.L. and T.J.T. acknowledge grants from the National Science and Technology Council of Taiwan 106-2119-M-007-021-MY3 and 109-2112-M-007-010-MY3. J.P.W. acknowledges support from NSF AST-2107841. H.-W.Y. acknowledges support from the National Science and Technology Council (NSTC) in Taiwan through the grant NSTC 110-2628-M-001-003-MY3 and from the Academia Sinica Career Development Award (AS-CDA-111-M03).


\bibliography{References}{}
\bibliographystyle{aasjournal}




\restartappendixnumbering

\begin{appendix}


\section{Additional continuum images}\label{ap:ContinuumImages}

\begin{figure*}
\centering
\includegraphics[width=\linewidth,trim={0cm 9cm 0cm 1.5cm},clip]{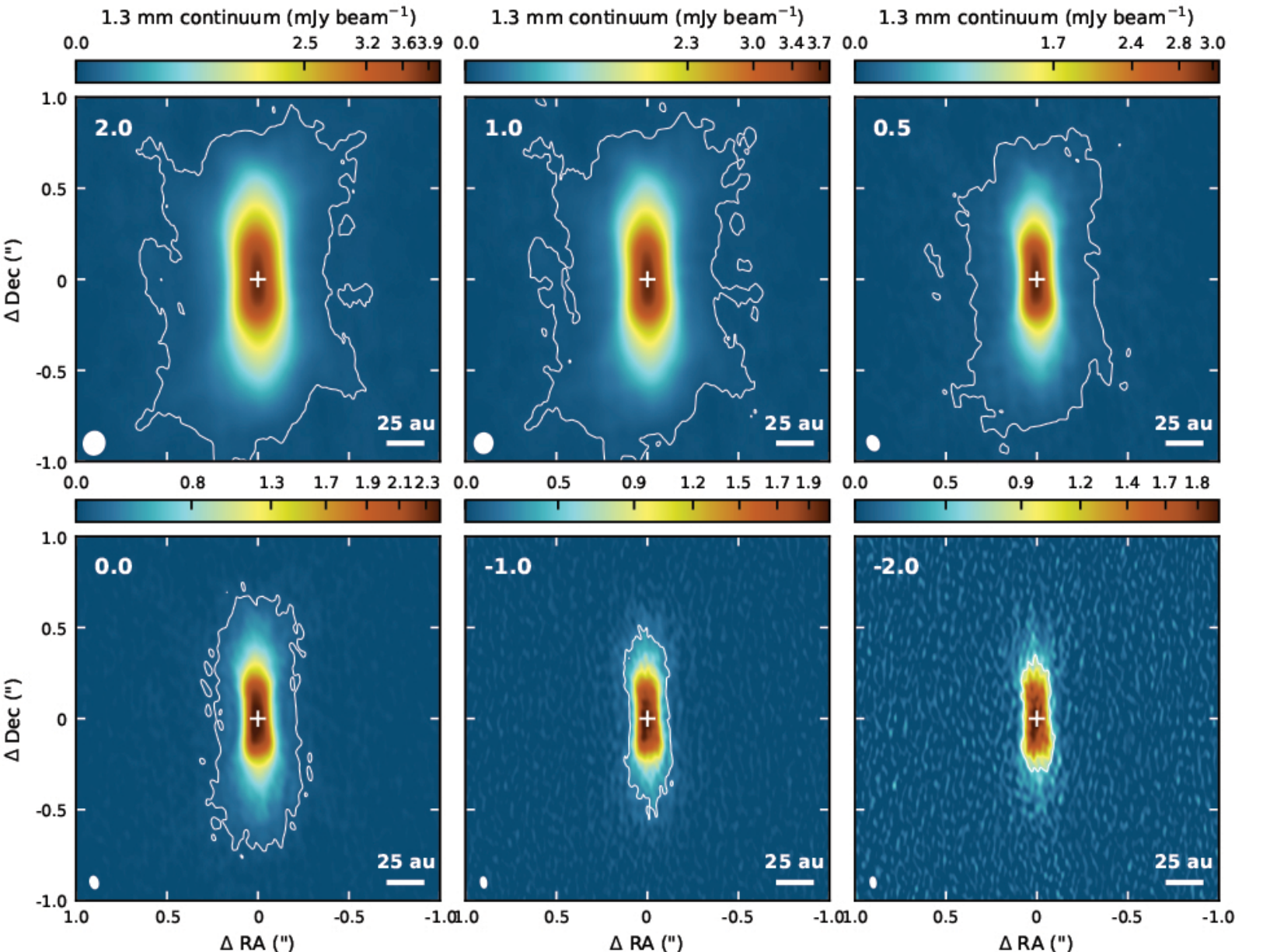}
\caption{ALMA 1.3 mm continuum image of L1527 imaged with different robust parameters as listed in the top left corner of each panel. The robust = $-$0.5 image is shown in the left panel of Fig.~\ref{fig:Continuum}. The beam size is depicted by the white ellipse in the bottom left corner of each panel. The solid white contour marks the 5$\sigma$ level and the white cross indicates the source position.}
\label{fig:ContinuumOverviewRobust}
\end{figure*}

\begin{figure*}
\centering
\includegraphics[width=\linewidth,trim={0cm 13.7cm 0cm 1.0cm},clip]{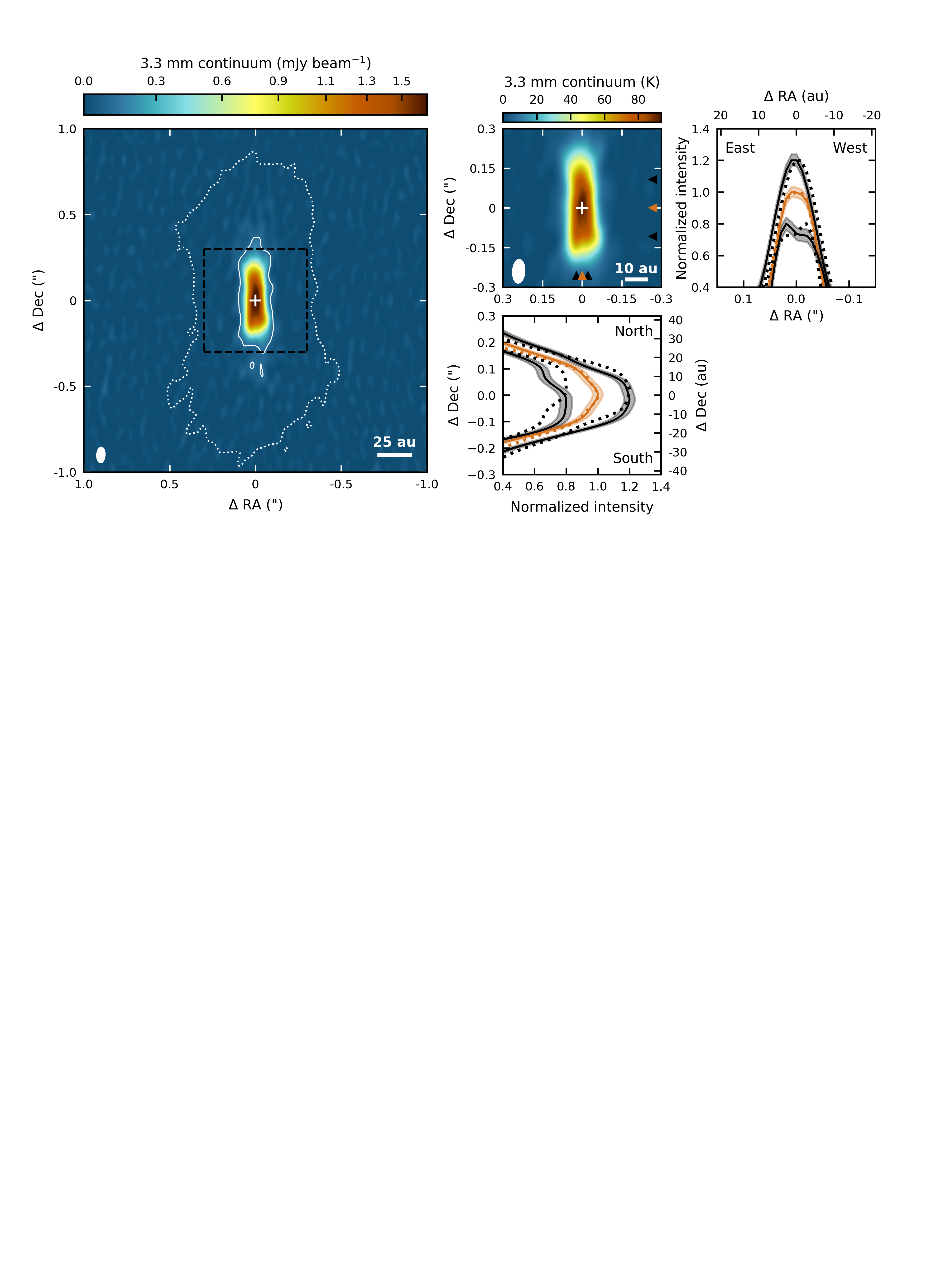}
\caption{ALMA 3.3 mm continuum image of L1527 from archival data created with the eDisk data reduction script. The left panel shows the full extent of the continuum imaged with a robust parameter of $-$1.0 (color scale; beam size of 0\farcs086 $\times$ 0\farcs043), with the solid white contour marking the 5$\sigma$ level (0.13 mJy beam$^{-1}$). The dotted white contour marks the 5$\sigma$ level (0.04 mJy beam$^{-1}$) of the map imaged with a robust parameter of 2.0 (beam size of 0\farcs225 $\times$ 0\farcs173). The black dashed square shows the region depicted in the top middle panel, where the brightness temperature of the continuum is displayed. The top right panel shows the normalized intensity along the disk minor axis at source position (orange) and at 15 au to the north (top black curve) and south (bottom black curve). This is a slightly different offset compared to what is shown for the 1.3 mm image, as the asymmetry is strongest at larger radii at 3.3 mm. The bottom right panel shows the normalized intensity along the disk major axis at source position (orange) and at 3 au to the east (left black curve) and west (right black curve). Black and orange triangles in the continuum image (top middle panel) mark the locations of the intensity profiles. The black curves are shifted by 0.2 in normalized intensity with respect to the orange curves for better visibility. The shaded region depicts the 3$\sigma$ level and the dotted lines are the mirror images of the solid lines to highlight the asymmetries. }
\label{fig:Continuum_3mm}
\end{figure*}

Figure~\ref{fig:ContinuumOverviewRobust} presents an overview of the 1.3 mm  continuum toward L1527 imaged with different robust parameters in addition to the image made with a robust value of $-$0.5 as presented in Fig.~\ref{fig:Continuum} (left panel). In addition, Fig.~\ref{fig:Continuum_3mm} presents a similar image for the 3.3 mm continuum image from archival ALMA data (2017.1.00509.S, PI: N. Sakai) as shown in Fig.~\ref{fig:Continuum} for 1.3 mm. 

\section{Additional molecular line images} 

\begin{figure*}
\centering
\includegraphics[trim={0cm 13.3cm 0cm 0.4cm},clip]{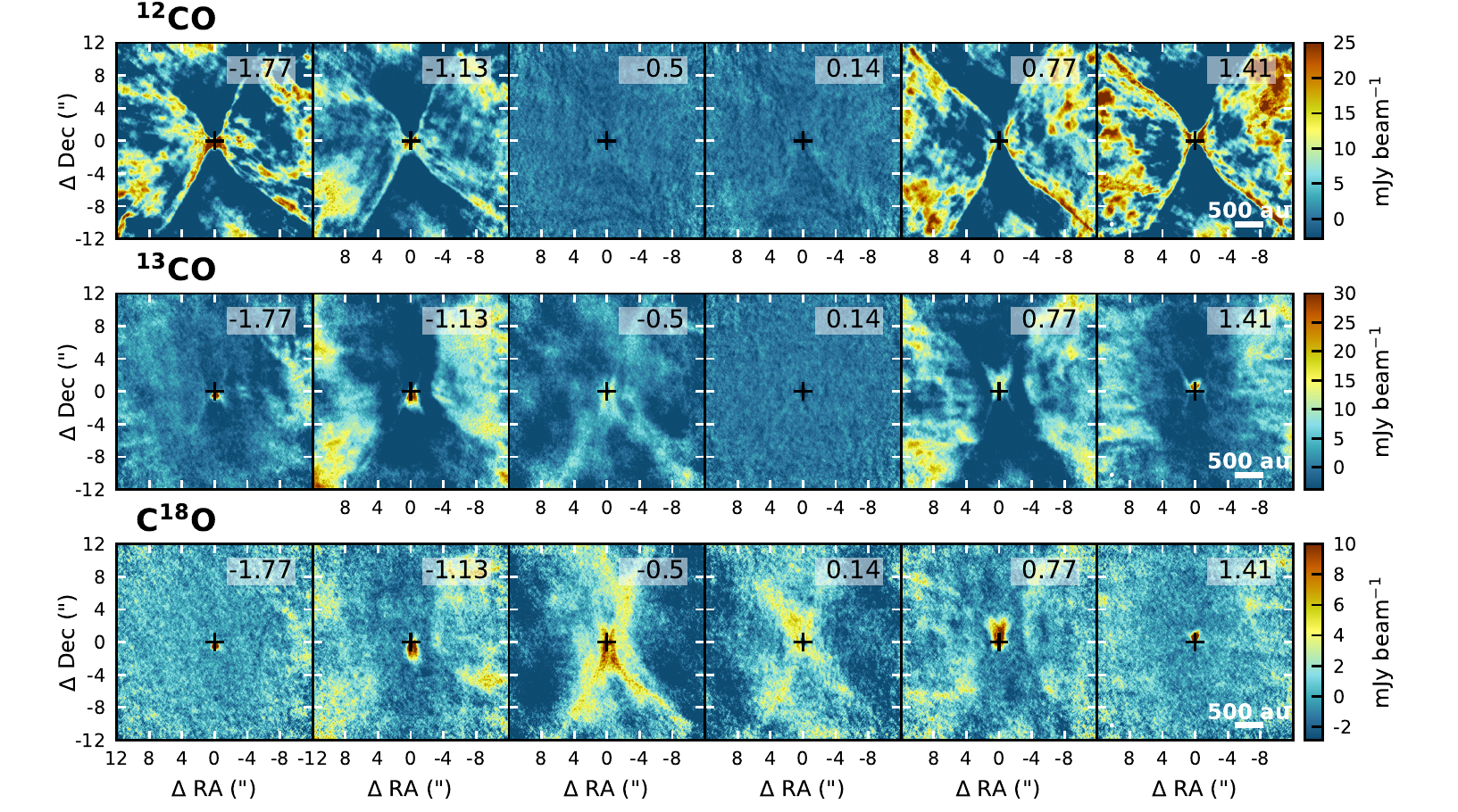}
\includegraphics[trim={0cm 13.3cm 0cm 0.4cm},clip]{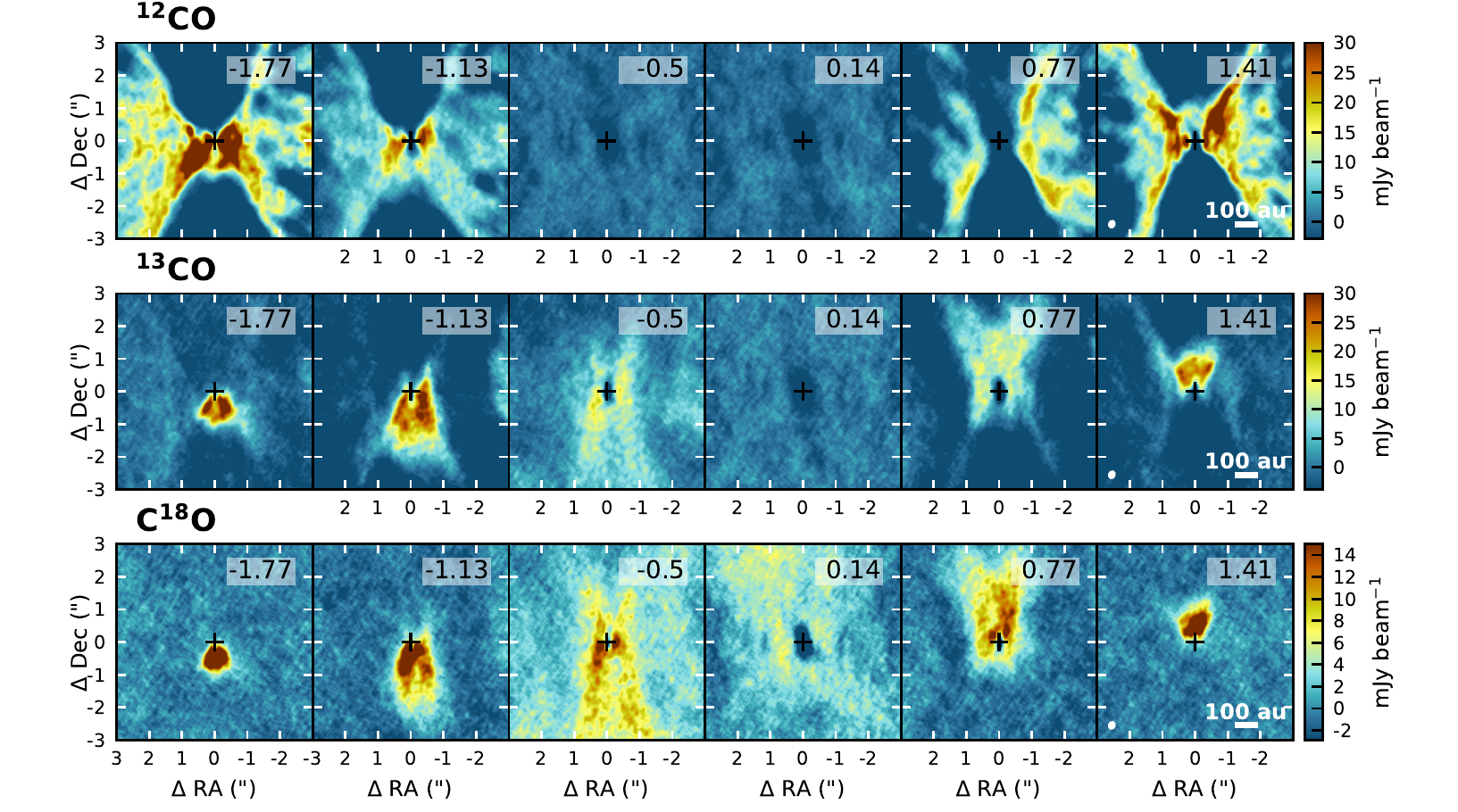}
\caption{Selected velocity channel maps of $^{12}$CO (first and fourth row), $^{13}$CO (second and fifth row) and C$^{18}$O (third and sixth row) as used for the RGB overlay in Fig.~\ref{fig:CO_channels_overlay}. The top three rows display the emission on scales of 24\asec (corresponding to the top row in Fig.~\ref{fig:CO_channels_overlay}) and the bottom three rows display the emission on scales of 6\asec (corresponding to the bottom row in Fig.~\ref{fig:CO_channels_overlay}).}
\label{fig:CO_channels}
\end{figure*}

\begin{figure*}
\centering
\includegraphics[trim={0cm 12.8cm 0cm 0.0cm},clip]{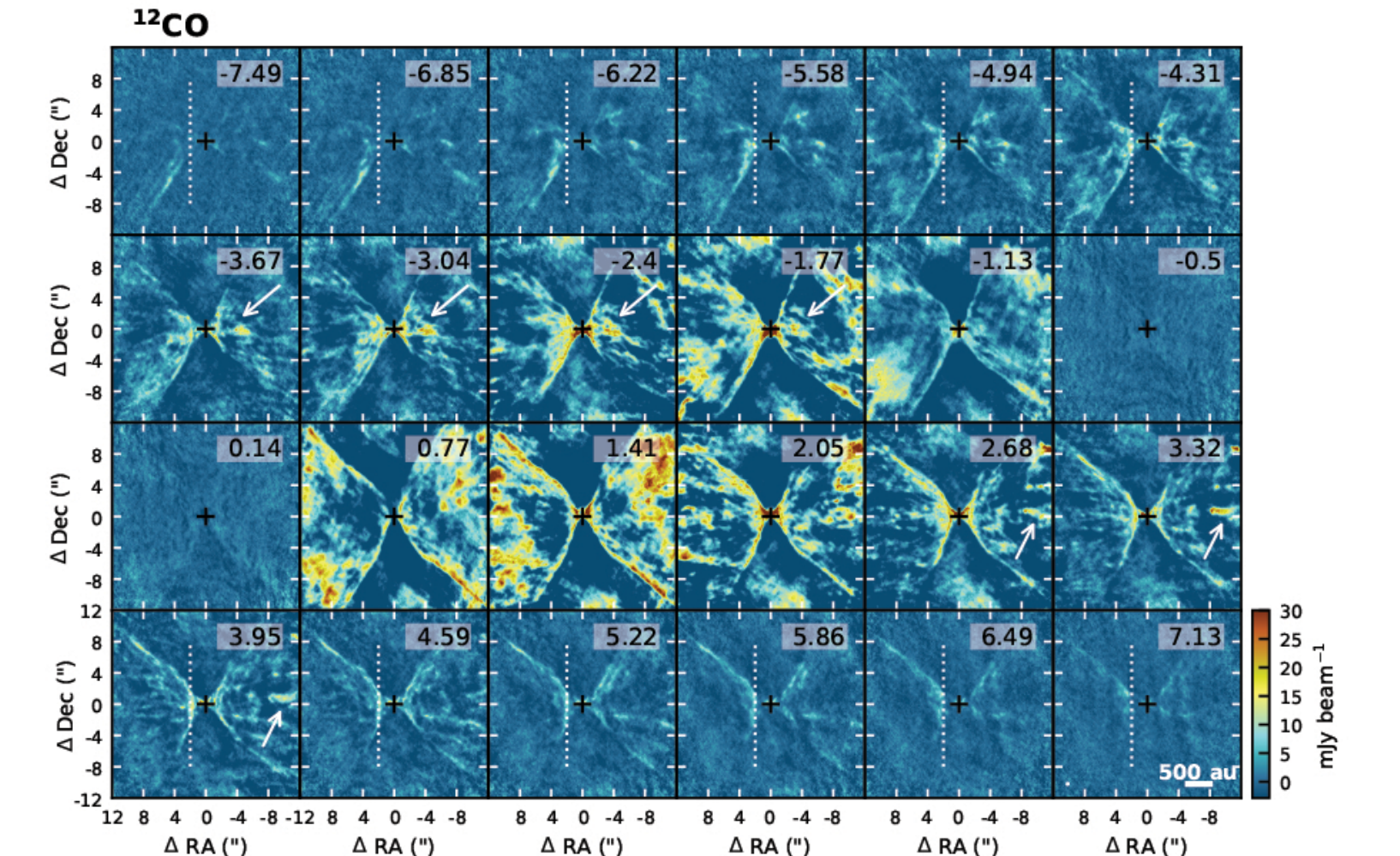}
\caption{Velocity channel maps of $^{12}$CO. The color scale is capped to highlight the large scale emission. A vertical white dotted line is drawn at the same position in the top and bottom panels to guide the eye with respect to the outward moving emission front at the highest velocity offsets. The white arrows at $-$3.67 -- $-$1.77 \kms and 2.68 -- 3.95 \kms highlight the potential jets. The beam is depicted in the bottom left corner of the bottom right panel, and the velocity with respect to the system velocity is listed in the top right corner of each panel. }
\label{fig:12CO_channels}
\end{figure*}

\begin{figure*}
\centering
\includegraphics[trim={0cm 12.8cm 0cm 0.0cm},clip]{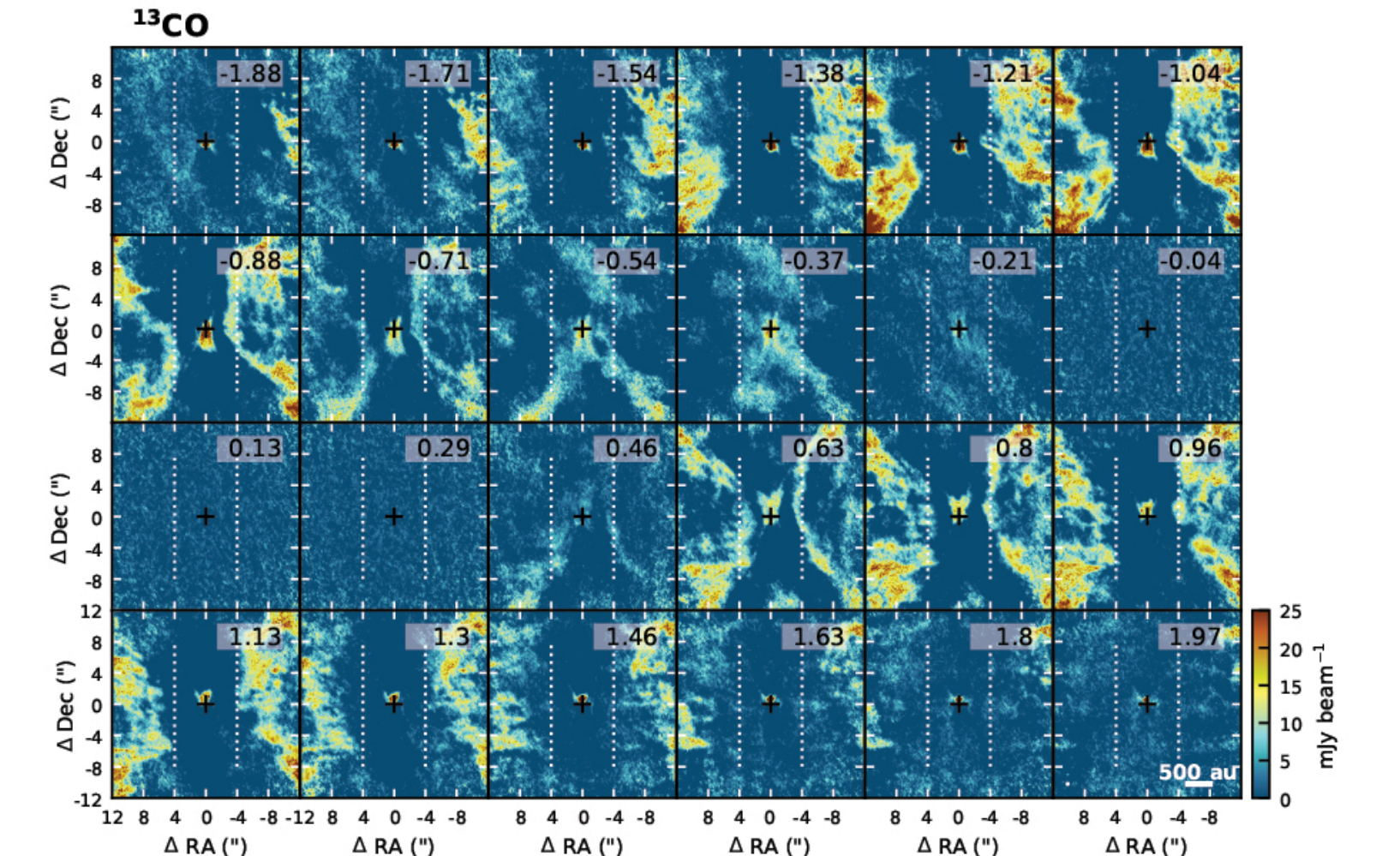}
\caption{Low and intermediate velocity channel maps of \thCO highlighting the emission in the outflow cavity. Two vertical white dotted lines are drawn at the same position in each panel to guide the eye with respect to the outward moving emission fronts with higher velocity offsets. The beam is depicted in the bottom left corner of the bottom right panel, and the velocity with respect to the system velocity is listed in the top right corner of each panel.}
\label{fig:13CO_channels}
\end{figure*}

\begin{figure*}
\centering
\includegraphics[trim={0cm 12.6cm 0cm 0.0cm},clip]{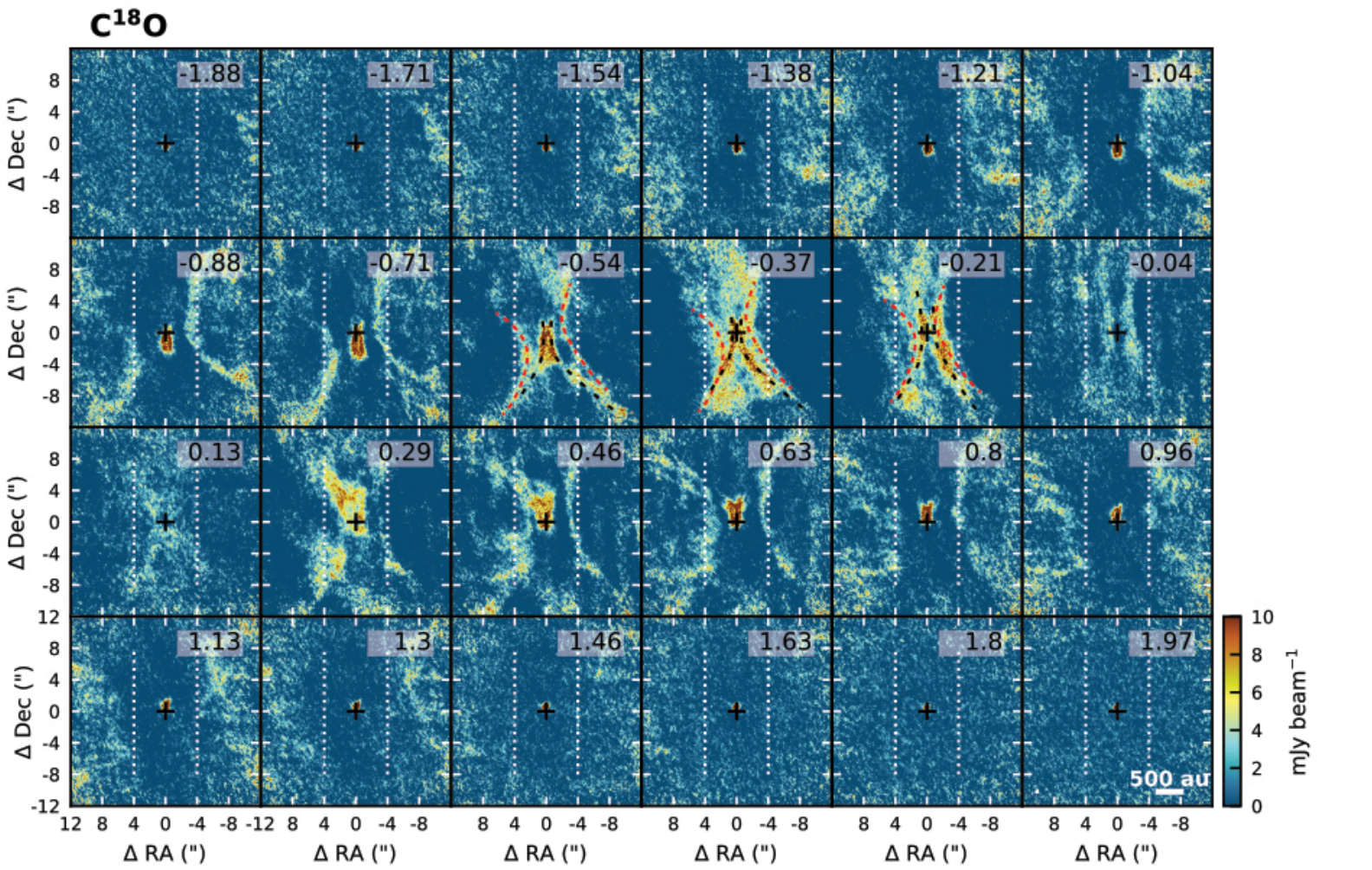}
\caption{As Fig.~\ref{fig:13CO_channels}, but for \CeO. Different components are outlined in three velocity channels ($-$0.54 -- $-$0.21 \kms), with dashed red lines indicating outflowing material and dashed black lines marking the envelope surface or cavity wall. These curves are drawn by hand and merely serve as a guide for the eyes.}
\label{fig:C18O_channels}
\end{figure*}

\begin{figure*}
\centering
\includegraphics[trim={0cm 19cm 0cm 0.0cm},clip]{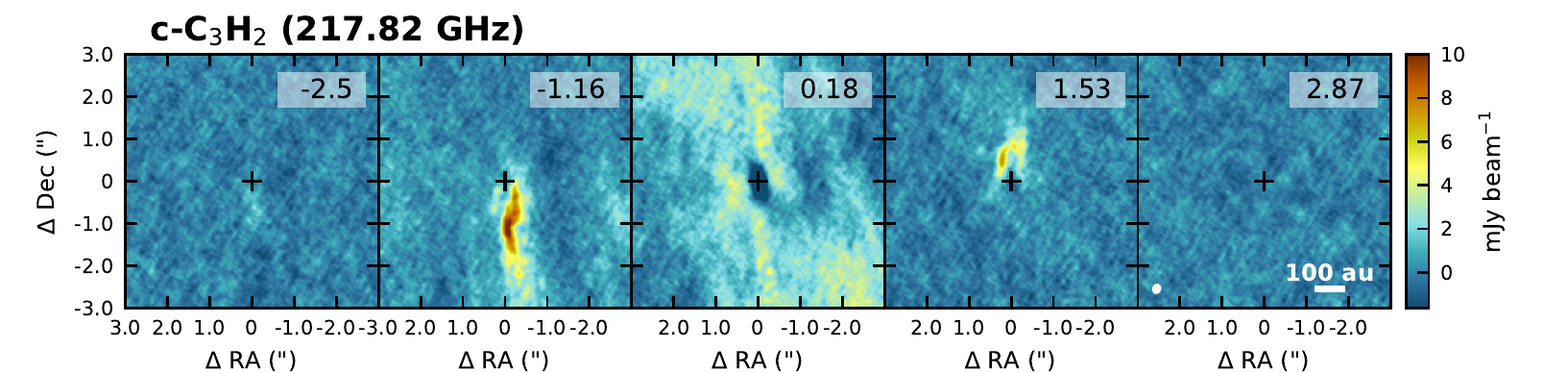}
\includegraphics[trim={0cm 19cm 0cm 0.0cm},clip]{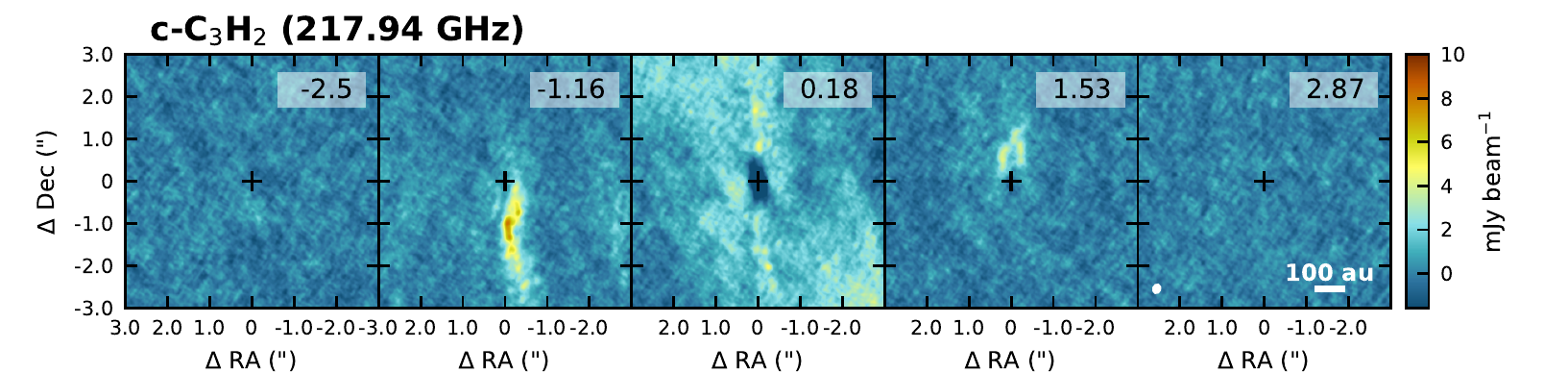}
\includegraphics[trim={0cm 19cm 0cm 0.0cm},clip]{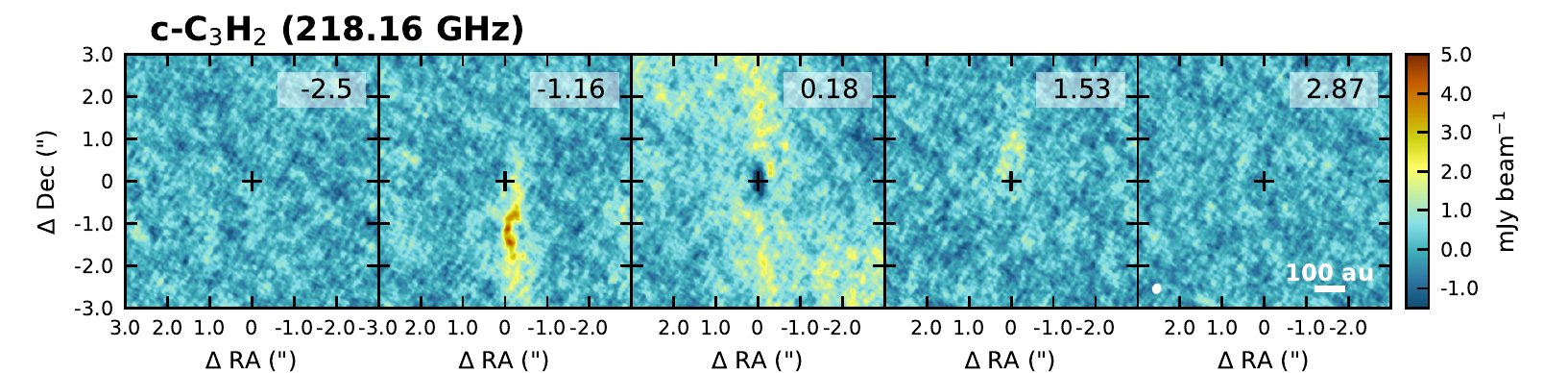}
\caption{Velocity channel maps of c-C$_3$H$_2$ $6_{0,6}-5_{1,5}$ blended with $6_{1,6}-5_{0,5}$ (top row), $5_{1,4}-4_{2,3}$ (middle row) and $5_{2,4}-4_{1,3}$ (bottom row).}
\label{fig:c-C3H2_channels}
\end{figure*}

\begin{figure*}
\centering
\includegraphics[trim={0cm 19cm 0cm 0.0cm},clip]{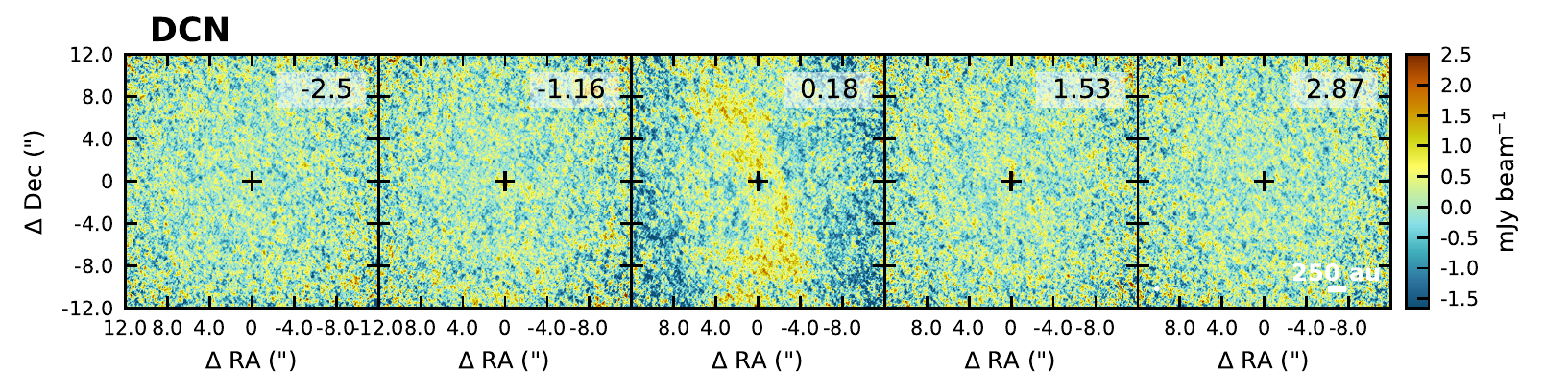}
\includegraphics[trim={0cm 19cm 0cm 0.0cm},clip]{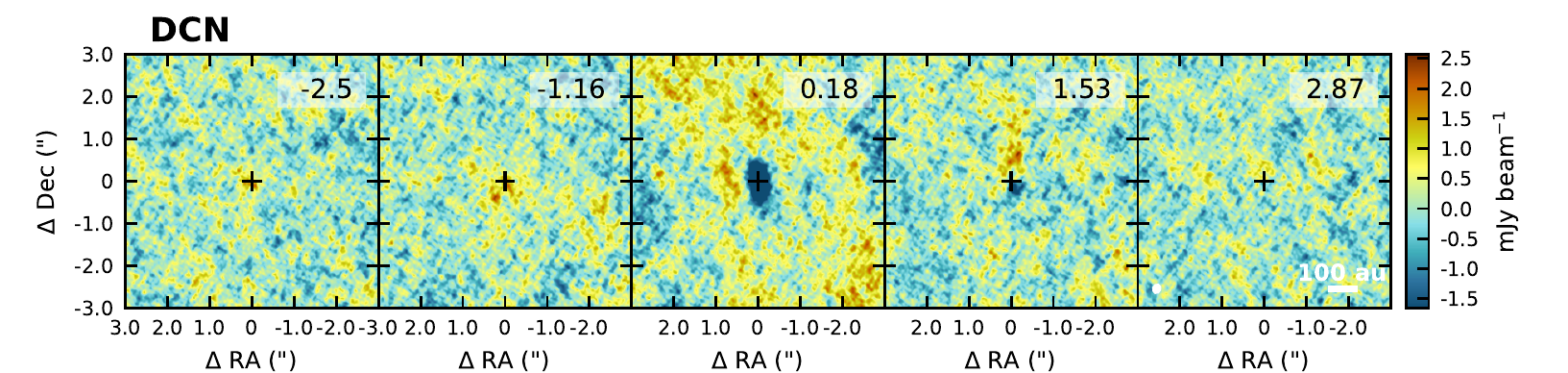}
\caption{Velocity channel maps of DCN on a 24\asec scale (top row) and 6\asec scale (bottom row).}
\label{fig:DCN_channels}
\end{figure*}

\begin{figure*}
\centering
\includegraphics[trim={0cm 13.2cm 0cm 0.0cm},clip]{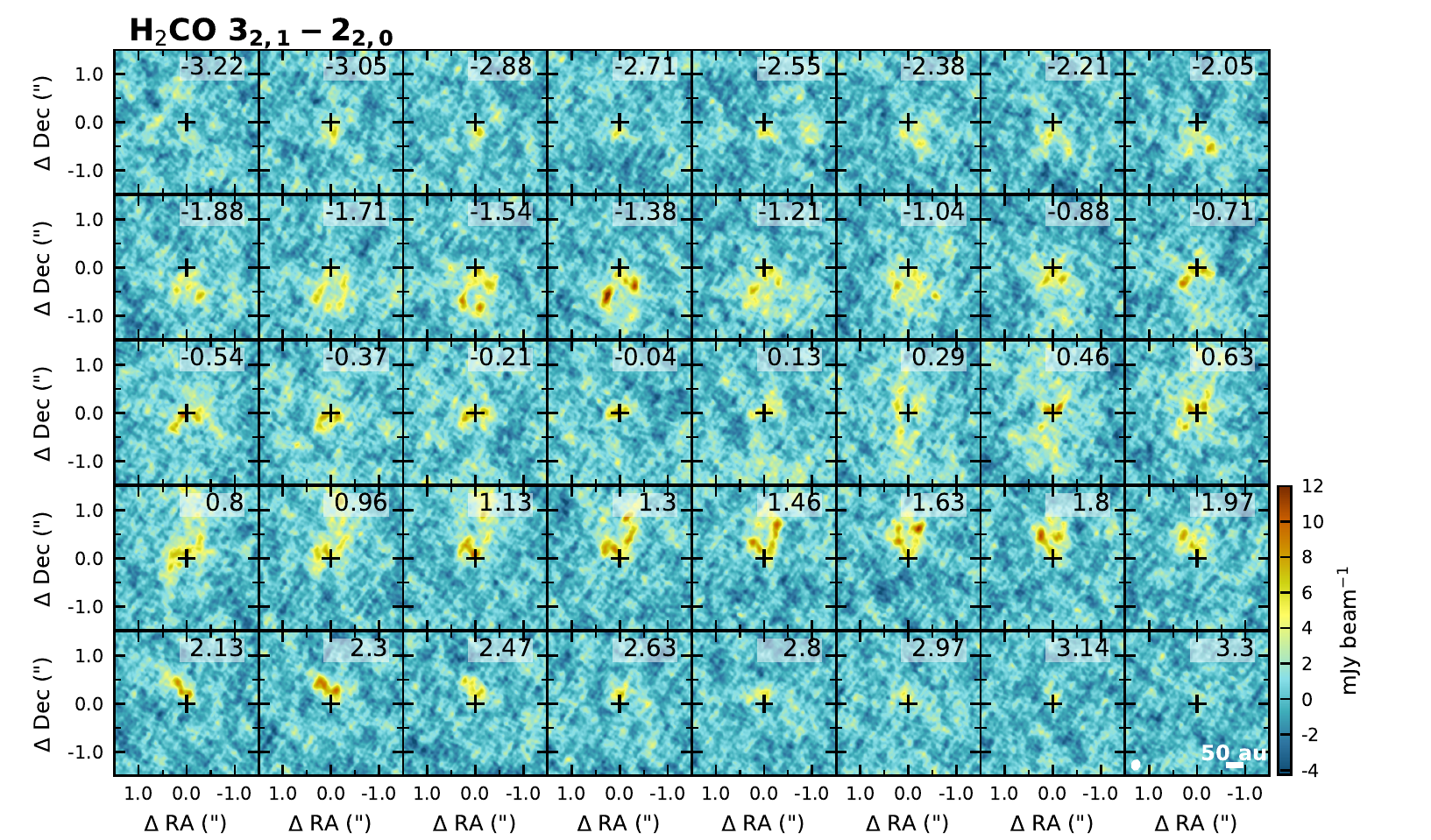}
\caption{Velocity channel maps of H$_2$CO $3_{2,1}-2_{2,0}$.}
\label{fig:H2CO_channels_full}
\end{figure*}

\begin{figure*}
\centering
\includegraphics[trim={0cm 11.5cm 0cm 0.0cm},clip]{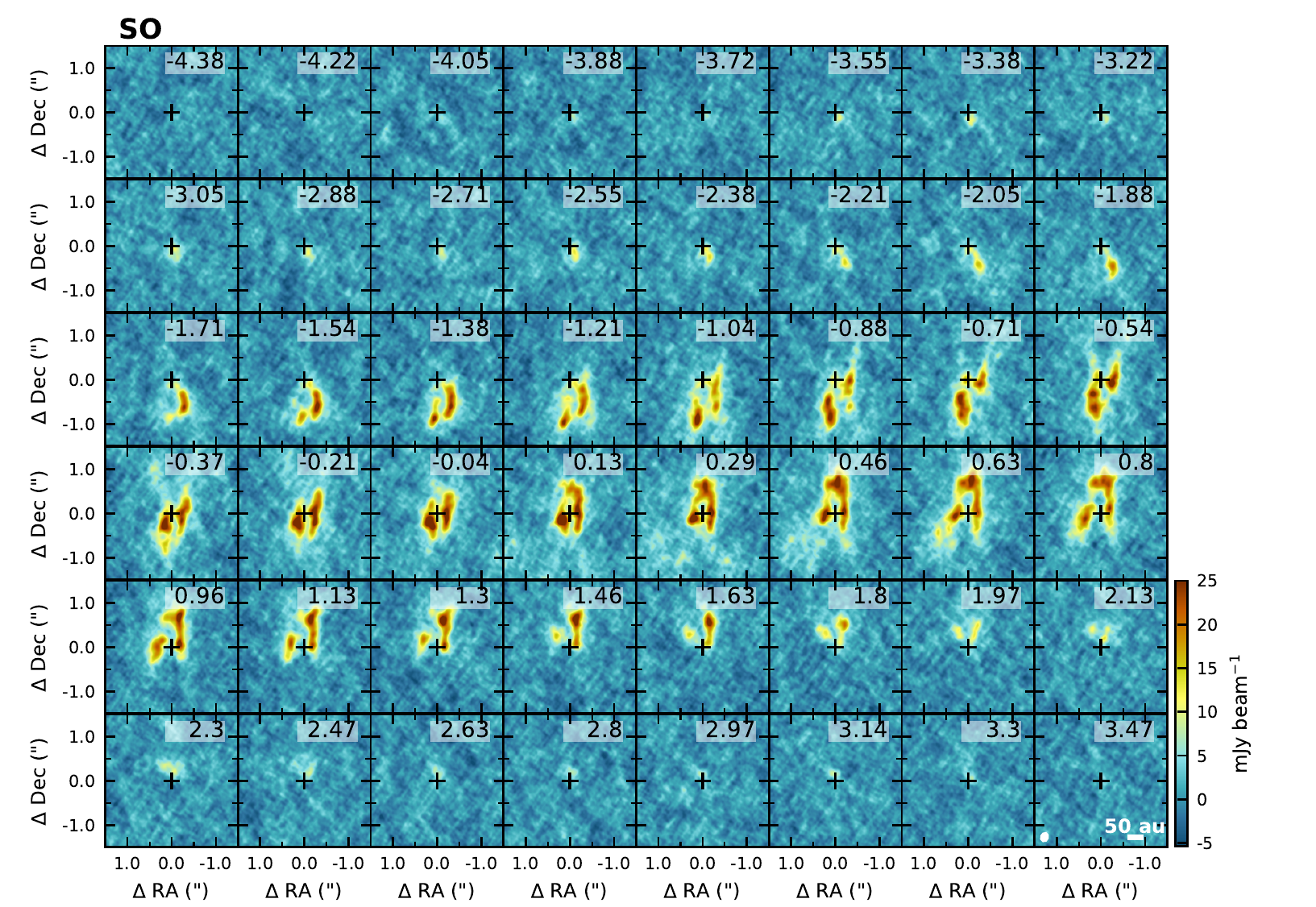}
\caption{Velocity channel maps of SO}
\label{fig:SO_channels_full}
\end{figure*}

\begin{figure}
\centering
\includegraphics[trim={0cm 8cm 0cm 9.5cm},clip]{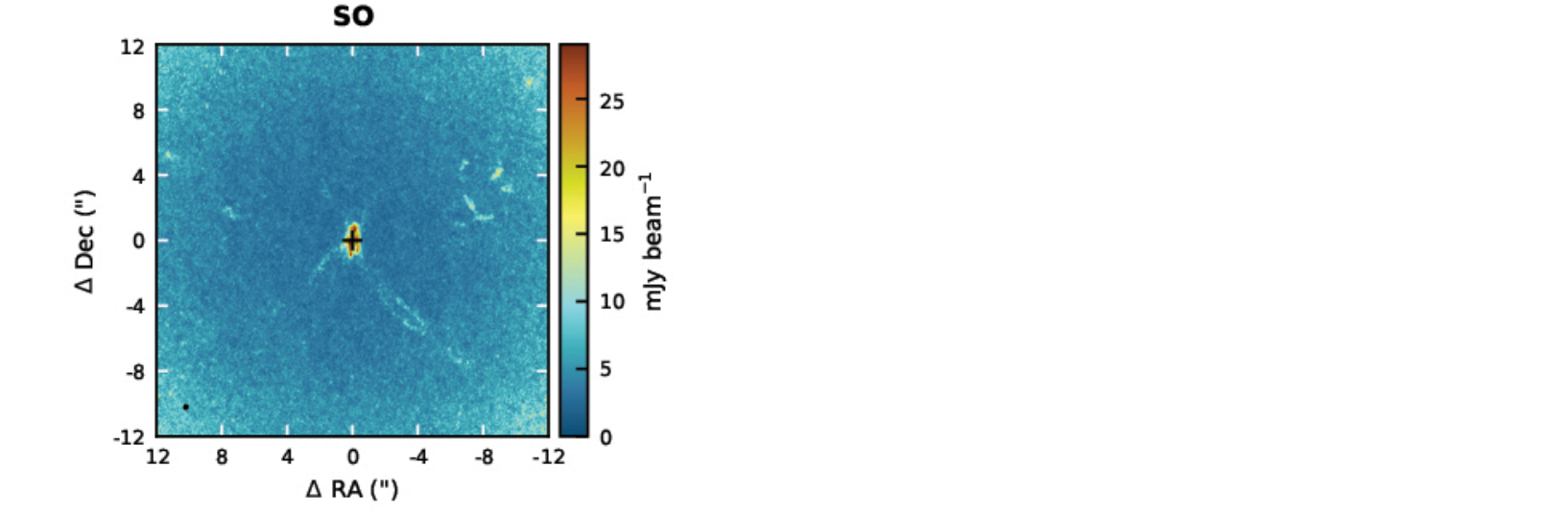}
\caption{Peak intensity (moment eight) map of SO. }
\label{fig:SO_M8}
\end{figure}

The individual \twCO, \thCO, and \CeO channel maps used to make the RGB overlay (Fig.~\ref{fig:CO_channels_overlay}) are shown in Fig.~\ref{fig:CO_channels}. In addition, Figs.~\ref{fig:12CO_channels}--\ref{fig:C18O_channels} present \twCO, \thCO, and \CeO channel maps to highlight the outflowing material visible in the three CO isotopologues. Velocity channel maps for the \cCtHt transitions are displayed in Fig.~\ref{fig:c-C3H2_channels}, for DCN in Fig.~\ref{fig:DCN_channels}, for \HtCO $3_{2,1}-2_{2,0}$ in Fig.~\ref{fig:H2CO_channels_full}, and for SO in Fig.~\ref{fig:SO_channels_full}. Figure~\ref{fig:SO_M8} shows the SO moment eight (peak intensity) map. 

\section{Additional figures}

\begin{figure}
\centering
\includegraphics[trim={0cm 16.2cm 7cm 0cm},clip]{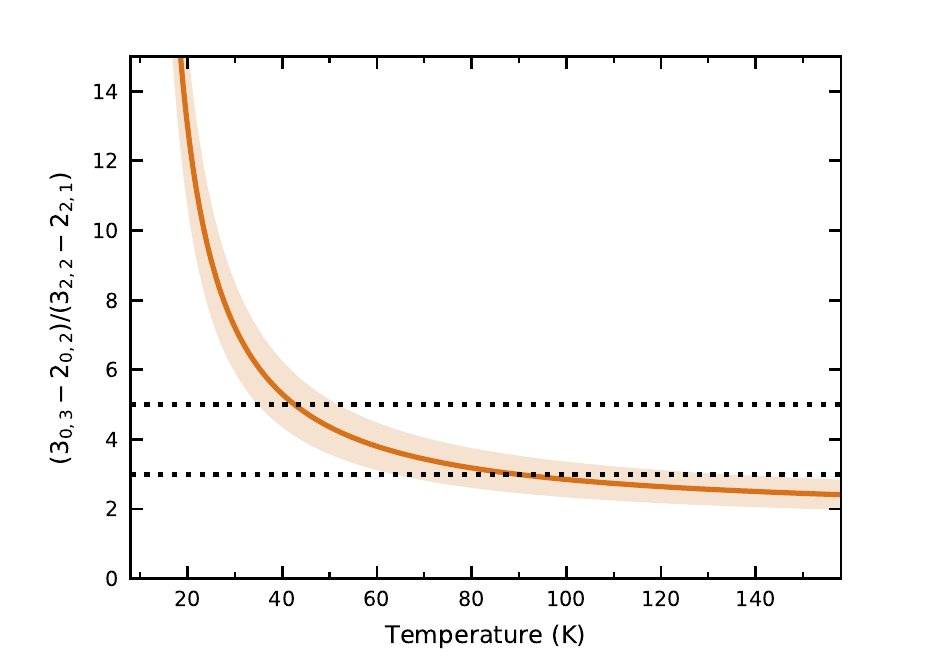}
\caption{Line ratio of the \HtCO $3_{0,3}-2_{0,2}$ and $3_{2,2}-2_{2,1}$ transitions as function of temperature for optically thin emission in LTE. The shaded area marks the uncertainty for a signal-to-noise ratio of 6 for the $3_{2,2}-2_{2,1}$ transition and 12 for the $3_{0,3}-2_{0,2}$ transition. Horizontal dotted lines marking line ratios of 3 and 5 are drawn for reference.}
\label{fig:H2CO_ratio}
\end{figure}

Figure~\ref{fig:H2CO_ratio} presents the \HtCO $3_{0,3}-2_{0,2}$/$3_{2,2}-2_{2,1}$ line ratio as function of temperature for optically thin emission in LTE.  

\end{appendix}

\end{document}